\nonstopmode

\newcommand{\im}{{\mathrm{Im}\,}}
\newcommand{\eV}{\U{eV}}
\newcommand{\mat}[1]{\hbox{\boldmath{$#1$}\unboldmath}}
\newcommand{\rad}{\U{rad}}
\newcommand{\bohr}{\, a_0}
\newcommand{\Cal}[1]{{\cal #1}}
\newcommand{\U}[1]{\,{\rm{#1}}}
\newcommand{\I}[1]{_{\mathrm{#1}}}
\newcommand{\imag}{{\rm i}}
\newcommand{\euler}{\mathrm e}
\newcommand{\Int}{\int\limits}
\newcommand{\Lim}{\lim\limits}
\newcommand{\differential}{\>\mathrm d}
\newcommand{\bra}[1]{\left<\right.\!#1\!\left.\right|}
\newcommand{\ket}[1]{\left|\right.\!#1\!\left.\right>}
\newcommand{\XUV}{\textsc{xuv}}
\newcommand{\IR}{\textsc{ir}}

\documentclass[pra,aps,twocolumn,showpacs,nopreprintnumbers]{revtex4}
\usepackage{bm,bbm,amsmath,amssymb,subeqnarray,graphicx}
\usepackage[nativepdf,bookmarks,bookmarksopen,bookmarksnumbered,raiselinks,breaklinks,%
pdftitle={Transient absorption and reshaping of ultrafast XUV light by laser-dressed helium},%
pdfauthor={Mette B. Gaarde, Christian Buth, Jennifer L. Tate, Kenneth J. Schafer},%
pdfsubject={Atomic physics},%
pdfkeywords={quantum optics, XUV, photoabsorption, Auger decay, %
light-matter interaction, helium, attosecond, electromagnetically induced %
transparency for x-rays, EIT, Autler-Townes doublet, control, Floquet, %
wavepacket propagation}]{hyperref}

\begin{document}
\title{Transient absorption and reshaping of ultrafast \XUV{}~light
by laser-dressed helium}
\author{Mette B. Gaarde$^{1,2}$}
\email{gaarde@phys.lsu.edu}
\author{Christian Buth$^{3,2,1}$}
\email{christian.buth@web.de}
\author{Jennifer L. Tate$^1$}
\author{Kenneth J. Schafer$^{1,2}$}
\affiliation{$^1$Department of Physics and Astronomy, Louisiana State
University, Baton Rouge, Louisiana~70803, USA}
\affiliation{$^2$The PULSE Institute for Ultrafast Energy Science,
SLAC National Accelerator Laboratory, Menlo Park, California 94025, USA}
\affiliation{$^3$Max-Planck-Institut f\"ur Kernphysik,
Saupfercheckweg~1, 69117~Heidelberg, Germany}
\date{\today}

\begin{abstract}
We present a theoretical study of transient absorption and reshaping
of extreme ultraviolet (\XUV) pulses by helium atoms dressed with a
moderately strong infrared (\IR) laser field.
We formulate the atomic
response using both the frequency-dependent absorption cross section
and a time-frequency approach based on the time-dependent dipole
induced by the light fields.
The latter approach can be used in cases when an ultrafast dressing
pulse induces transient effects, and/or when the atom exchanges energy with multiple frequency components of the \XUV\  field.
We first characterize the dressed atom response by calculating the frequency-dependent absorption cross section for \XUV\ energies between 20 and 24 eV for several dressing wavelengths between 400 and 2000 nm and intensities up to 10$^{12}$ W/cm$^2$. We find that for dressing wavelengths near 1600 nm, there is an Autler-Townes splitting of the $1s \to 2p$~transition that can potentially lead to transparency for absorption of \XUV{}~light tuned to this transition. We study the effect of this \XUV\ transparency in a macroscopic helium gas by
incorporating the time-frequency approach into a solution of the coupled Maxwell-Schr\"odinger equations. We find rich temporal reshaping dynamics when a 61~fs \XUV\ pulse resonant with the $1s \to 2p$ transition propagates through a helium gas dressed by an 11~fs, 1600~nm laser pulse.
\end{abstract}

%
%
%
%
%
%

\preprint{arXiv:1010.4558}
\maketitle

\renewcommand{\onlinecite}[1]{\cite{#1}}

\section{Introduction}
The advent of ultrafast  \XUV\ and even x-ray light sources that can be synchronized to optical or \IR\ laser pulses has given rise to several recent studies of the transient absorption of such radiation by laser-dressed atoms, both experimental~\cite{Johnsson:AC-07, Glover:CX-09, Ranitovic2010013008, Goulielmakis:RO-10, Mauritsson2010053001} and theoretical~\cite{Wickenhauser:TR-05, Buth:ET-07, Pfeifer200811}. For example, many of the experiments done in attosecond physics involve the transient absorption of attosecond \XUV\ radiation by atoms interacting with an \IR\ laser field. This is because the strong-field process of high harmonic generation (\textsc{hhg}), which is used to produce the attosecond \XUV\ radiation as either single pulses or trains of pulses, results in the \XUV\ field being precisely synchronized with the driving \IR\ field \cite{Drescher:TR-02, Johnsson:AC-07}. It is then possible to perform experiments using the \XUV\ field and a replica of the original \IR\ field with attosecond precision \cite{Remetter10323}. Glover {\it et al.} also showed that it is possible to overlap pulses of synchrotron-produced soft x-ray radiation with an 800 nm dressing laser in a study of laser-induced transparency in neon~\cite{Glover:CX-09}.

In this paper we
explore how an ultrafast \XUV\ pulse interacts with a simple atom, helium, in the presence of a moderately strong \IR\ field which may be either shorter or longer in duration than the \XUV\ pulse. We have as our goal formulating theoretical methods that can be used to calculate the absorption and emission of \XUV\ radiation by strongly dressed atoms even when the \XUV\ pulses are on the femtosecond time scale, and may include multiple frequencies in a comb.  In addition we want to be able to study the transient absorption and reshaping of radiation as it propagates through a macroscopic amount of gas. We will restrict ourselves in this study to cases where the \IR\ laser dresses the atom without appreciably exciting it, leaving higher \IR\ intensities for a future paper. We will also restrict ourselves to \XUV\ wavelengths and intensities where single excitations below the first ionization threshold at 24.6 eV dominate the \XUV\ absorption. Even given these restrictions,  the \IR\ laser has a substantial impact on the Rydberg and continuum
states of the atom and, in this way, enables profound control
over resonant \XUV{} absorption~\cite{Boller:ET-91,Fleischhauer:ET-05,Buth:ET-07,%
Buth:TA-09,Swoboda:PM-10}.

In the calculations we present we will consider the simplest case, where there are just two radiation fields, one which dresses the atom and one which is absorbed and possibly reshaped.
The fundamental problem of a two-color field like this has been studied before in the context of x-ray absorption by
neon~\cite{Buth:ET-07,Santra:SF-07,Buth:RA-10,Young:US-10,Glover:CX-09},
argon~\cite{Buth:AR-08}, and krypton~\cite{Buth:TX-07} atoms.
Specifically, the examination of laser-dressed atoms led to the discovery of
electromagnetically induced transparency~(\textsc{eit}) for x-rays~\cite{Buth:ET-07},
better characterized as Autler-Townes splitting~\cite{Fleischhauer:ET-05} because the
transparency is not predominantly caused by destructive interference.
There have also been several studies of helium in the context of the two-color problem we are discussing. It was investigated with an
optical laser and the \XUV{}~free electron laser in
Hamburg~\cite{Laarmann:PH-05,Charalambidis:CPH-06,Laarmann:RCPH-06}, and
the impact of laser-dressing helium
on the production of \XUV\ radiation via high harmonic generation~(\textsc{hhg}) was studied theoretically in~\cite{Cormier:ED-97}.

We begin by characterizing the single atom response in terms of the cross section for absorption of \XUV\ radiation of frequency $\omega\I{X}$. First we  calculate the linear, frequency-dependent \XUV\ absorption cross section using a Floquet-like method (non-Hermitian perturbation theory -- \textsc{nhpt}) that treats the \XUV\ field as a monochromatic source. This method has been extensively tested in the context of  x-ray absorption on the $1s \to 3p$~resonance of laser-dressed neon, alluded to above~\cite{Buth:ET-07, Buth:TX-07}.
Next, we outline a method using direct integration of the time-dependent
Schr\"odinger equation (\textsc{tdse}) which achieves essentially the same goal using pulses of finite duration. The cross section is extracted by projecting out the initial state from the final state wave function. The two methods, \textsc{nhpt} and the \textsc{tdse}-projection method, are shown to agree when the \XUV\ pulse bandwidth is very small. The \textsc{tdse}-projection approach is, however, potentially more flexible in dealing with situations where the dressing laser couples many states of the atom. We find that the \textsc{tdse}-projection method can be used with a reasonable amount of effort to study laser-dressed absorption over a wide range of \XUV\ frequencies and dressing wavelengths. As an example of the usefulness of the method, we show representative results for several dressing wavelengths between 0.4 and 2.0 $\mu m$.

We next extend the treatment of the \XUV\ interaction to deal with cases where the atomic response varies as a function of time or frequency in a non-trivial way. This could for instance be because the \IR\ dressing pulse is so short that non-adiabatic effects cause the cross section to vary substantially over the bandwidth of the pulse. Another interesting situation is when the \IR\ pulse is so strong that multiphoton processes cause the atom to exchange energy with the light field over a large range of frequencies in many different orders of nonlinearity, so that it is no longer practical to (artificially) separate the linear/non-linear absorption from the driven linear/non-linear emission. We therefore develop a time-frequency approach to the atomic response, based on the time-dependent energy exchange between the atom and the light fields.  Our method is similar in spirit to the treatment provided by for instance Tannor \cite{Tannor07} and Pollard and Mathies \cite{Pollard1992} with the important difference that we are not separating the atomic response into different linear and non-linear orders but keeping everything in one frequency dependent response function.
We find that when we use long ($\sim$30 fs) \XUV\ pulses we get good agreement between linear absorption cross sections calculated using the time-frequency and the \textsc{tdse}-projection cross section. Having obtained this good agreement over a range of frequencies and dressing laser intensities gives us confidence that we can calculate the full time-frequency response of the dressed atom.

Finally we show how this time-frequency approach is consistent with our solution of the coupled Maxwell wave equation (\textsc{mwe}) and the \textsc{tdse}. This allows for a generalized, {\it ab-initio} description of linear and non-linear absorption, emission, and phase matching in a macroscopic medium. We apply this formalism to studying the propagation of an \XUV\ pulse in a macroscopic helium gas dressed by a moderately intense 11~fs, 1600~nm laser pulse. We find that the \XUV\ pulse, which is resonant with the $1s \to 2p$~transition in the undressed atom, undergoes rich temporal  absorption and reshaping dynamics.

The paper is structured as follows.
In Sec.~\ref{sec:theory}, we first discuss the three formalisms
for calculating absorption cross sections of laser-dressed atoms. In Sec.~\ref{Macroscopic} we present our framework for the macroscopic calculations.
Then we use the methods to study laser-dressed helium;
computational details are given in Sec.~\ref{sec:compdet} and
results are presented in Sec.~\ref{sec:results}.
We end the paper with a brief conclusion in Sec.~\ref{sec:conclusion}.

\section{Single atom response}
\label{sec:theory}

This section contains three derivations of the one-photon absorption cross
section for \XUV\ light interacting with an atom in the presence of a long wavelength dressing field.
All three formalisms are based on the single active electron (\textsc{sae}) approximation, and in all cases we use linearly polarized fields where the \IR\ and \XUV\ polarization vectors are parallel. We use atomic units through-out this section~\cite{Szabo:MQC-89}.

\subsection{Non-Hermitian Rayleigh-Schr\"odinger perturbation theory}
\label{sec:rspt}

Our \textsc{nhpt} treatment of dressed \XUV\ absorption is discussed in detail in references~\cite{Buth:NH-04,Buth:TX-07,Buth:AR-08}. Here we provide a brief account to highlight the essential steps in the derivation and to facilitate a discussion of the other two formalisms.

In the \textsc{nhpt} formalism,  the  one-photon \XUV{}~absorption cross section follows from
\begin{equation}
 \label{eq:xsect}
 \sigma = 2 \, \frac{\Gamma\I{I}}{J_{\mathrm X}} \; ,
\end{equation}
where $J_{\mathrm X}$~is the constant
\XUV{}~photon flux of a continuous wave \XUV\ light source~\cite{Buth:TX-07}, and $\Gamma\I{I}$ is
the transition rate from the initial state to Rydberg orbitals or the
continuum. The factor of~2 accounts for the number of electrons in the atomic
orbital which is used as the initial state~$\ket{\mathrm I}$.

To determine~$\Gamma\I{I}$ with \textsc{nhpt}
the full Hamiltonian of an atom in two-color light~$\hat H = \hat H_0
+ \hat H_1$ is decomposed into a strongly interacting part~$\hat H_0
= \hat H\I{AT} + \hat H\I{EM,L} + \hat H\I{I,L} + \hat H\I{EM,X}$
that contains the atomic electronic structure~$\hat H\I{AT}$ in
Hartree-Fock-Slater approximation~\cite{Slater:AS-51,Slater:XA-72}.
The interaction with light is expressed in terms of nonrelativistic quantum
electrodynamics~\cite{Buth:NH-04,Buth:TX-07,Buth:AR-08};
the free \IR\ laser and \XUV{}~fields are~$\hat H\I{EM,L}$ and
$\hat H\I{EM,X}$, respectively, and the interaction of the atomic
electrons with the laser field is~$\hat H\I{I,L}$.
The weak interaction with the \XUV{}~light is represented
by~$\hat H_1 = \hat H\I{I,X}$~\cite{Buth:TX-07}.

Next we represent~$\hat H_0$ in a complex-symmetric direct product basis of
electronic states---without the initial state~$\ket{\mathrm I}$---and photonic number states.
In doing so, we assume that the initial state and its energy~$E\I{I}$
are not noticeably influenced by the laser dressing.
The matrix representation of~$\hat H_0$,
\begin{equation}
 \label{eq:diag_H0}
 \mat H_0^{(m)} \vec c_F^{\>(m)} = E_F^{(m)} \vec c_F^{\>(m)} \; ,
\end{equation}
is diagonalized, yielding eigenvectors~$\vec c_F^{\>(m)}$ which represent the expansion
coefficients of new laser-dressed states~$\ket{F^{(m)}}$ for
eigenvalues~$E_F^{(m)}$~\cite{Buth:TX-07}.
Here, $m$~is the magnetic quantum number that is conserved for linearly
polarized light.

When the Hamiltonian $\hat H_0$ is represented in the new basis of laser-dressed states [Eq.~(\ref{eq:diag_H0})],
the excitation or ionization of a ground-state electron of an atom
due to \XUV{}~photoabsorption is described as a resonance in the spectrum of
the non-Hermitian, complex-symmetric representation of the
Hamiltonian in the basis ~$\{ \ket{\mathrm I} , \ket{F^{(m)}} | \, \forall F,m
\}$~\cite{Kukulin:TR-89,Moiseyev:CS-98,Santra:NH-02}.
The complex energy of the resonance state that $\ket{\mathrm I}$
becomes due to the coupling to excited states and the continuum via
\XUV{}~light is usually refered to as the Siegert energy~\cite{Kukulin:TR-89,Siegert:DF-39} and satisfies
\begin{equation}
 \label{eq:siegert}
 E\I{res} = E\I{R} - \imag \, \Gamma\I{I} / 2 \; .
\end{equation}
The real part of the resonance energy is~$E\I{R}$, and $\Gamma\I{I}$ stands
for the transition rate from the ground state to a
laser-dressed Rydberg orbital or the laser-dressed continuum
[Eq.~(\ref{eq:diag_H0})] via photoabsorption.
We determine the Siegert energy [Eq.~(\ref{eq:siegert})] of the initial state $\ket{\mathrm I}$ in second-order non-Hermitian perturbation theory. The total transition rate out of $\ket{\mathrm I}$ is given by
\begin{equation}
 \label{eq:G_I_td}
 \Gamma\I{I} = 2 \  \im \biggl[ \sum_{m,F}
   \frac{\bra{\mathrm I} \hat H_1 \ket{F^{(m)}} \bra{F^{(m)}} \hat H_1
   \ket{\mathrm I}}{E_F^{(m)} - E\I{I}} \biggr] \;
\end{equation}
and the absorption cross section is finally obtained from Eqs.~(\ref{eq:G_I_td})
and (\ref{eq:xsect}) as:
\begin{equation}
 \label{eq:sigcomb}
 \sigma(\omega_{\mathrm X}) =
   8 \pi \, \alpha \, \omega_{\mathrm X} \  \im \biggl[ \sum_{m,F}
   \frac{(\mathcal{D}_F^{(m)})^2}{E_F^{(m)} - E\I{I} - \omega_{\mathrm X}}
   \biggr] \; .
\end{equation}
Here $\alpha$~denotes the fine-structure constant and
$\mathcal{D}_F^{(m)}$~is a complex-scaled transition
dipole matrix element between the initial
state $\ket{\mathrm I}$ and the $F$th laser-dressed atomic state
with projection quantum number~$m$~\cite{Buth:TX-07}.

\subsection{Projection treatment of \XUV\ absorption}
\label{sec:tdpt}

As an alternative to the treatment above, we can obtain the linear absorption cross section by a direct solution of the \textsc{tdse}
in the \textsc{sae} approximation~\cite{Schafer:NM-08}. The cross section is extracted by projecting the final state wave function obtained at the end of a finite pulse onto
the initial wave function. As such, we avoid calculating the dressed states directly, making explicit use of only the laser-free initial and final states.

To simplify the treatment of finite duration pulses when using the
projection method, we replace the quantum electrodynamic treatment
of \XUV\ radiation  in Sec.~\ref{sec:rspt}, by a semiclassical description of
light~\cite{Craig:MQ-84}.
We begin by choosing the vector potential of the \XUV{}~light of carrier frequency~$\omega\I{X}$ to be
\begin{equation}
\label{eq:xuvpulse_a}
  \vec {\cal A}\I{X}(t) = -\frac{{\Cal E}\I{X}(t)}{\omega\I{X}} \,
  \sin(\omega\I{X} \, t) \, \vec e_{\rm x}\; .
\end{equation}
The electric field of the \XUV{}~light field is then given by derivative with respect to time,  $\vec {\cal E}\I{X}(t)= -\partial  \vec {\cal A}\I{X}(t) /\partial t$:
\begin{equation}
\label{eq:xuvpulse}
  \vec {\cal E}\I{X}(t) = \left[{\Cal E}\I{X}(t)  \,
  \cos(\omega\I{X} \, t) +\frac{1}{\omega\I{X}}\frac{\partial {\cal E}\I{X}(t)}{\partial t}\sin(\omega\I{X}\,t)\right] \vec e_{\rm x}\; .
\end{equation}
Here, ${\cal E}\I{X}(t) = \sqrt{8 \pi \alpha I\I{X}(t)}$ is the
envelope of the \XUV\ pulse
and $I\I{X}(t)$ is its cycle-averaged intensity. Our specification of the vector potential in Eq.~\ref{eq:xuvpulse_a} ensures that
the integrated electric field and the vector potential at the end of the pulse, ${\cal A}(t_f)$, are zero when
${\cal E}\I{X}(t)$ is zero at the initial and final times. It leads to the second term on the right hand side of Eq.~\ref{eq:xuvpulse} which is
a small correction of order $\Delta\omega\I{X}/\omega\I{X}$ near the center of the pulse, for pulses with a bandwidth of
$\Delta \omega\I{X}$. By ensuring that ${\cal A}(t_f)=0$ we obtain results  which are independent of the electromagnetic gauge.

For a  Gaussian envelope pulse with a full width at half maximum (\textsc{fwhm}) duration of~$\tau\I{X}$,
the bandwidth of the pulse is given by $\Delta \omega\I{X} = 4 \ln 2/\tau\I{X}$. In our calculations, we
first specify $\Delta \omega\I{X}$ and this dictates the value of $\tau\I{X}$. The Gaussian envelope is
then approximated by a trigonometric pulse~\cite{Barth:TP-09}:
\begin{equation}
  \label{eq:GaussEnv}
  I\I{X}(t) = I\I{X,0} \> \cos^{2n} \Bigl( \frac{\pi t}{T_n}
    \Bigr) \; \theta \Bigl(\frac{T_n}{2} - |t| \Bigr) \equiv
    I\I{X,0} \>g_n(t) \; ,
\end{equation}
with an integer~$n>0$ and the Heaviside $\theta$~function~\cite{Arfken:MM-05}.
The total pulse duration is defined as
\begin{equation}
\label{eq:PulseDuration}
T_n = \frac{\pi \, \tau\I{X}}{2 \arccos 2^{-\frac{1}{2n}}} \; ,
\end{equation}
The envelope~(\ref{eq:GaussEnv}) converges rapidly to a Gaussian function in the limit~$\Lim_{n \to \infty} g_n(t) \to  \exp\left(-4 \ln 2
(\frac{t}{\tau\I{X}})^2\right)$.
Using the approximative function~(\ref{eq:GaussEnv}) instead of a true Gaussian function has the advantage
that it goes to zero on a finite support, which allows us to satisfy the requirement ${\cal A}(t_f)=0$ exactly.

In the \textsc{tdse}-projection formalism we also need to dress the atom with a laser field with frequency $\omega_{\mathrm L}$. We do this by using a laser field of the form
\begin{equation}
\label{eq:itpulse}
  \vec {\cal E}\I{L}(t) = {\cal E}\I{L}(t) \, \vec e\I{L}  \,
  \sin(\omega\I{L}  t) \; .
\end{equation}
The envelope function ${\cal E}\I{L}(t)$ is now a trapezoidal pulse with a linear ramp of one optical cycle at each end and a flat section that completely spans over the \XUV{}~pulse.  The pulse contains an integer number of laser cycles, so we again obtain zero vector potential at the end of the dressing pulse. We assume this field is too weak to excite or ionize the atom on its own, an assumption that we can explicitly check by running the calculation once without the \XUV{}~field.

To calculate the cross section for absorption we begin with the atom in its ground state $|\psi_I\rangle$ at time $t_0$ and use the grid-based methods of reference~\cite{Schafer:NM-08} to propagate the wave function forward in time until  the end of the combined \XUV\ and dressing pulse at time $t_f$. At this time we calculate the probability that the atom has remained in its ground state $P_I(t_f)$ by projecting the final wave packet $|\psi(t_f)\rangle$ onto the initial wave packet:
\begin{equation}
\label{eq:project}
P_I(t_f)=\left| \langle \psi_I  | \psi(t_f)\rangle \right|^2.
 \end{equation}
Given~$P\I{I}(\infty)=P\I{I}(t_f)$ from the \textsc{tdse} calculation, we obtain the probability that the atom is excited or ionized from~$1 - P\I{I}(t_f)$. Because we are dealing with a one-photon absorption process where we assume the intensity is well below saturation, a linear relation between \XUV{}~absorption rate and \XUV{}~photon flux holds:
$\Gamma(t) = \sigma(\omega_X) \, J\I{X}(\omega_X, t)$.
We use this assumption to transform the probability to absorb an \XUV{}~photon into an expression for the cross section:
\begin{equation}
\label{eq:excross}
2[1 - P\I{I}(\infty)] = \sigma(\omega_X)  \Int_{-\infty}^{\infty}
  J\I{X}(\omega_X,t) \differential t \; .
\end{equation}
This is equivalent to the steady state expression in Eq.~(\ref{eq:xsect}):
The factor~2 again stems from the two electrons in the spatial orbital~I which contribute equally.

The underlying assumption in Eq.~(\ref{eq:excross}) is that we can calculate the absorption cross section $\sigma(\omega_{\mathrm X})$ for a small range of frequencies $\Delta\omega_{\mathrm X}$ around $\omega_{\mathrm X}$ by calculating the response of the atom to a pulse of bandwidth $\Delta\omega_{\mathrm X}$.
For Eq.~(\ref{eq:excross}) to be meaningful, the cross section needs
to be approximately constant over the bandwidth~$\omega\I{X}$ of the pulse.
For a low-bandwidth pulse, we can further use the relation~$I\I{X}(t) \approx
\omega\I{X} \, J\I{X}(\omega_X,t)$ between photon flux and intensity.
Then, the time-integral on the left-hand side can be solved
analytically for the pulse shape~(\ref{eq:xuvpulse}).
In this way, we find the \XUV{}~absorption cross sections~$\sigma(\omega_X)$
from Eq.~(\ref{eq:excross}) by dividing
the probability to excite an atom out of the ground state~$1 - P\I{I}(t_f)$
by the integral over the \XUV{}~flux.

As we stated in the introduction, though we expect that the two methods for calculating the frequency-dependent absorption cross section should agree, the \textsc{tdse} projection approach is potentially more flexible in dealing with situations where the dressing laser couples many states of the atom which forces the Hamiltonian matrix in Eq.~\ref{eq:diag_H0} to be very large.

\subsection{Time-frequency treatment of ultrafast \XUV\ absorption}
\label{sec:timefreq}

In this section we extend the treatment of the \XUV\ interaction to deal with cases where the atomic response varies as a function of time or frequency in a non-trivial way. This could be because the dressing \IR\ pulse is substantially shorter than the \XUV\ pulse, or when non-linear interactions would cause the atom to exchange energy with multiple \XUV\ frequency components in different non-linear orders.

We start by deriving a frequency-dependent response function $\widetilde{S}(\omega)$ from the time-dependent energy exchange between the atom and the light field. $\widetilde{S}(\omega)$ is defined so that when integrated over all frequencies, it yields the total excitation probability. This includes excitation to continuum states, {\it i.e.}, ionization. We can then express the total energy gained by the atom from the light fields, $\Delta E$, as the sum over the frequency dependent excitation probability $\widetilde{S}(\omega)$ times the photon energy:
\begin{eqnarray}
  \Delta E  = \int_{-\infty}^{\infty}  \omega\, \widetilde{S}(\omega)\,d\omega .
\end{eqnarray}

To calculate the response function we use that the total atomic energy gain can also be expressed as a sum over the rate at which energy is gained:
\begin{eqnarray}
\Delta E  = \int_{-\infty}^{\infty} \omega\, \widetilde{S}(\omega)\,d\omega = \int_{-\infty}^{\infty} \frac{dE}{dt}\, dt.
\label{SfromDeltaE}
\end{eqnarray}
We calculate this rate directly from our one electron Hamiltonian, $H=H_A+{\cal E}(t)\,z$, as:
\begin{eqnarray}
\frac{dE}{dt}=\frac{d}{dt}\langle \psi | H | \psi\rangle =
\langle \psi | \frac{\partial H}{\partial t} | \psi \rangle
= \langle z \rangle\frac{\partial{\cal E}}{\partial t}.
\end{eqnarray}
We note that ${\cal E}(t)$ is the full electric field consisting of the sum of the dressing laser and the \XUV\ fields. This means that we are simultaneously treating the exchange of energy between the atom and all frequencies of the light field.
In the following we will denote  $\langle z\rangle(t)$ by $z(t)$. The time-dependent dipole moment is related to $z(t)$ by $d(t) = -z(t)$ for a single electron.
We now calculate $\Delta E$:
\begin{eqnarray}
\Delta E&=&\int_{-\infty}^{\infty} z(t) \frac{\partial{\cal E}}{\partial t}\,dt\\[1em]
&=& -\int_{0}^{\infty}\omega\,  2\, {\rm Im}\!\left\{\tilde{z}(\omega)\widetilde{\cal E}^*(\omega)\right\}\, d\omega \; .
\end{eqnarray}
In this derivation we have used that both $z(t)$ and ${\cal E}(t)$ are real functions of time so that $\tilde z(-\omega) = \tilde z^*(\omega)$ and $\widetilde{\cal E}(-\omega) = \widetilde{\cal E}^*(\omega)$.
Using Eq.~(\ref{SfromDeltaE}) we then have an expression for the response function:
\begin{equation}
\widetilde{S}_+(\omega) = -2\, {\rm Im}\!\left\{\tilde{z}(\omega)\widetilde{\cal E}^*(\omega)\right\} \hspace{2 em} \omega > 0,
\label{eq:respfunction}
\end{equation}
where the $+$ subscript on $\widetilde{S}_+(\omega)$ explicitly indicates that we are only integrating over positive frequencies.

We calculate the dipole spectrum in the \textsc{sae} approximation $\tilde d_{SAE}(\omega)$ via the time-dependent acceleration $a(t)$:
\begin{equation}
a(t) = \frac{d^2 z}{dt^2} = -  \langle \psi(t) | [H,[H,z]] |\psi(t) \rangle ,
\label{eq:acceleration}
\end{equation}
The dipole spectrum is then given by $\tilde d_{SAE}(\omega) = \tilde a(\omega)/\omega^2$, where $\tilde a(\omega)$ denotes the Fourier transform of $a(t)$.  The full (two-electron) dipole moment is  $\tilde{d}(\omega) = 2\tilde d_{SAE}(\omega)$.

In the weak-\IR\ limit where it is meaningful to talk about an absorption cross section, we can write the frequency-dependent energy exchange function $\omega \widetilde S(\omega)$ by means of a generalized cross section $\tilde \sigma(\omega)$ and the spectral energy density of the electric field, $ \omega {\widetilde J}(\omega)$. The spectral flux ${\widetilde J}(\omega)$ is defined as~\footnote{Our Fourier transformation convention
is~${\cal E}(t) = \frac{1}{\sqrt{2\pi}} \Int_{-\infty}^{\infty} \widetilde{\cal E}(\omega) \, \euler^{-i\omega t}\, \differential \omega$ and
$\widetilde{\cal E}(\omega)=\frac{1}{\sqrt{2\pi}} \Int_{-\infty}^{\infty} {\cal E}(t)\, \euler^{i\omega t}\, \differential t$.
}:
\begin{eqnarray}
{\widetilde J}(\omega)=\frac{1}{4\pi \alpha \omega} \left|\widetilde{\cal E}(\omega)\right|^2,
\end{eqnarray}
This means that once we calculate the response function $\widetilde S(\omega)$, the generalized cross section is given by:
\begin{equation}
\tilde \sigma(\omega)= \frac{4\pi \alpha \omega\, \widetilde{S}(\omega)}{|\widetilde{\cal E}(\omega)|^2}.
\end{equation}
 Inserting the response function from Eq.~(\ref{eq:respfunction})  we obtain the cross section, now defined for both positive and negative frequency components:
\begin{equation}
\sigma(\omega)
= 8\pi \alpha \omega \,{\rm Im}
\!\left\{\frac{\tilde{d}_{SAE}(\omega)}{\widetilde{\cal E}(\omega)}\right\}.
\label{sigmaTD}
\end{equation}
This equation is the generalized, time-frequency, multi-mode equivalent of Eq.~(\ref{eq:xsect}) which was derived for the steady-state case.

To calculate the generalized cross section in Eq.~(\ref{sigmaTD}), and the macroscopic polarization field described in the following sub-section, we multiply the time-dependent acceleration in Eq.~(\ref{eq:acceleration}) with a window function $W(t)$, $a_W(t) =  a(t) W(t)$ and calculate $\tilde d_{SAE}(\omega)$ from the Fourier transform of $a_W(t)$. In Eq.~(\ref{sigmaTD}) we also calculate $\widetilde{\cal E}(\omega)$ from $W(t) {\cal E}(t)$ for normalization purposes. The window function on the time-dependent acceleration is necessary in particular in those cases where the \XUV\ light is resonant with an atomic transition. The \XUV\ light then induces a strong coherence between the ground state and the excited state which in the numerical calculation will go on ``ringing'' until long after the \XUV\ pulse is over. This ringing does not correspond to stimulated emission or absorption of \XUV\ radiation.
The window function we use is a trigonometric function as given in Eq.~(\ref{eq:GaussEnv}) and is in general chosen to have the same \textsc{fwhm} duration as the longer of the \IR\ and \XUV\ pulses. The choice of window function has some influence on the value of the cross section for the un-dressed atom around the field-free resonances. When the atom is laser-dressed so that the \XUV\ light is no longer absorbed as strongly, the ringing is strongly suppressed by the laser field and the influence of the window function is very small.

It is interesting to note here that for intense or few-cycle \IR\ fields, and/or  for multi-mode \XUV\ fields, the sign of the response function $\widetilde{S}_+(\omega)$ (and therefore the sign of the generalized cross section) for a particular frequency $\omega$ in Eq.~(\ref{eq:respfunction}) can be positive or negative.  When $\widetilde{S}_+(\omega)$ is positive the atom will predominantly absorb light of that frequency, and when $\widetilde{S}_+(\omega)$ is negative the atom will predominantly emit light of that frequency. This makes the response function a powerful tool for studying the dynamics of the light-atom energy exchange, in particular in combination with a sliding time-window on the time-dependent acceleration. This would in principle allow for the time-resolution of when different frequencies are absorbed or emitted during a dynamical process. We will discuss a simple application of this in connection with the macroscopic reshaping of an \XUV\ pulse presented in the Results section.

\section{Macroscopic response, including absorption}
\label{Macroscopic}

As we will show at the end of this section, the relationship derived in the previous section, between the dipole spectrum driven by an arbitrary pulse and the absorption cross section for the frequencies contained in that pulse, is consistent with our general framework for the interaction between an ultrafast, multi-color pulse and a macroscopic medium. This framework consists of the coupled solutions of the \textsc{mwe} and the \textsc{tdse} for all frequencies $\omega$ of the electric field $\tilde E(\omega)$ of the multi-color pulse. We will express all quantities in SI units in this section. In a frame that moves at the speed of light, and in the slowly evolving wave approximation  which works well even for few-femtosecond pulses~\cite{Brabec:IF-00}, the \textsc{mwe} takes the following form:
\begin{equation}
\nabla^2_{\perp}\tilde{\cal E}(\omega)+\frac{2i\omega}{c}\frac{\partial \tilde {\cal E}(\omega)}{\partial z}=-\frac{\omega^2}{\epsilon_0 c^2}(\tilde P(\omega)+\tilde P_{ion}(\omega)).
\label{MWE}
\end{equation}
The electric field $\tilde {\cal E}(\omega)$ and the source terms $\tilde P(\omega)$ and $\tilde P_{ion}(\omega)$ are also functions of the cylindrical coordinates $r$ and $z$.
We solve this equation by space-marching through the helium gas, at each plane $z$ in the propagation direction calculating the response terms $\tilde P(\omega)$ and $\tilde P_{ion}(\omega)$ via numerical integration of the \textsc{tdse}, and then using them to propagate to the next plane in $z$. The macroscopic polarization field $\tilde P(\omega)$ is calculated from two times the one-electron single atom dipole moment $\tilde d_{SAE}(\omega)$:
\begin{equation}
\tilde P(\omega) =  2 \rho \tilde d_{SAE}(\omega) = \frac{2\rho e }{\omega^2 \sqrt{2\pi}} \int_{-\infty}^{\infty} a(t)W(t) \euler^{i\omega t}\, \differential t,
\end{equation}
where $\rho$ is the atomic density, $a(t)$ is the time-dependent acceleration calculated as described in Sec.~\ref{sec:timefreq}. As the driving field for the \textsc{tdse} calculation we use  the evolving electric field ${\cal E}(t)$ at the plane $z$. This means that $\tilde P(\omega)$ in general includes both the linear and nonlinear response of the atom to the multicolor field. The term $\tilde P_{ion}(\omega)$  is due to the space- and time-dependent free-electron contribution to the refractive index and is also calculated within the \textsc{sae-tdse}, see \cite{Gaarde:MA-08}. This term is very small in the cases considered in this paper and we will ignore it hereafter.

By calculating the source terms in each $z$-plane and using them to propagate to the next $z$-plane, we are coupling both the linear and non-linear response generated in one step back into the full electric field so that it can contribute to the driving electric field in the next step.  In much of the work described in the literature, see for instance \cite{L'Hu922778, Priori2000, Brabec:IF-00, Yak0715351, Gaarde:MA-08}, the non-linear response is separated from the linear response, and the propagation of the newly generated radiation (via nonlinear processes) is separated from the propagation of the driving field. Absorption and dispersion of different frequency components of the light fields are then added separately, typically using tabulated, frequency-dependent values.
It has been shown in a number of papers that such an approach offers a very complete description of both the generation of new frequencies via nonlinear processes, and the macroscopic effects of phase matching and ionization-driven reshaping of the ultrafast propagating pulse  \cite{L'Hu922778, Priori2000, Brabec:IF-00, Yak0715351, Gaarde:MA-08}. However, it cannot describe ultrafast or dynamical reshaping of the \XUV\ pulses driven by for instance absorption, dispersion, or laser-induced transparency. More generally, processes that are due to the combined response to the strong dressing or driving laser field and the weaker \XUV\ fields are not described in a self-consistent manner because the generated radiation is not included into the driving field.

In the following we will argue that the approach presented in this paper, which allows us to calculate the non-linear response of the dressed atom, also allows us to describe the absorption and dispersion of the ultrafast pulses in a self-consistent manner, to within the \textsc{sae} approximation.
Let us first rewrite the macroscopic polarization field as:
\begin{equation}
\tilde P(\omega) = \rho \tilde d(\omega) = \rho [{\rm Re}(\frac{\tilde d(\omega)}{\tilde {\cal E}(\omega)})+i {\rm Im}(\frac{\tilde d(\omega)}{\tilde {\cal E}(\omega)})] \tilde {\cal E}(\omega)
\end{equation}
The last term on the right is proportional to the generalized cross section in Eq.~(\ref{sigmaTD}). By inserting this expression into the  MWE in Eq.~(\ref{MWE}) we get:
\begin{equation}
\nabla^2_{\perp}\tilde {\cal E}(\omega)+2i\frac{\partial \tilde {\cal E}(\omega)}{\partial z}=
         -\frac{\omega}{\epsilon_0 c}\rho {\rm Re}(\frac{\tilde d(\omega)}{\tilde {\cal E}(\omega)})\tilde {\cal E}(\omega)
         -i\rho \tilde \sigma(\omega) \tilde {\cal E}(\omega).
\label{MWEsigma}
\end{equation}
The second term on the right hand side clearly will lead to absorption at frequency $\omega$ with absorption coefficient $\rho \tilde \sigma(\omega)$ when $\tilde \sigma(\omega)$ is positive, which is the case in the weak field limit when the atomic response is linear. In this linear case, the first term on the right hand side can likewise be interpreted as a generalized expression for the dispersion experienced in the gas medium, with the frequency dependent correction to the refractive index given by $\Delta \tilde n(\omega) = \frac{\rho}{2\epsilon_0}{\rm Re}(\frac{\tilde d(\omega)}{\tilde {\cal E}(\omega)})$.
The strength of our time-dependent approach is that even when the driving field is strong enough to induce non-linear processes,
we are able to treat all of the linear and non-linear processes within one time-dependent calculation, rather than artificially separating processes of different non-linearities and assigning them a frequency- and intensity-dependent weight.

\section{Computational details}
\label{sec:compdet}

Computations with the time-independent theory of Sec.~\ref{sec:rspt}
were carried out with the \textsc{dreyd} computer program from the
\textsc{fella} suite~\cite{fella:pgm-V1.3.0}.
The computational parameters are specified in analogy to
Ref.~\onlinecite{Buth:TX-07}.
However, in this work, we do \emph{not} rely on the Hartree-Fock-Slater
mean-field approximation~\cite{Slater:AS-51,Slater:XA-72} to describe
the atomic electronic structure.
Instead, we use  a pseudopotential for helium, constructed from the ground state Hartree-Fock potential, calculated on a very fine radial grid by standard iterative methods~\cite{Koonin98}. We set the $K$~edge of helium to the value of~$E\I{1s}
= -24.5786 \eV$. Next, the radial Schr\"odinger equation is solved with the pseudopotential
where the solution, the radial part of the atomic orbitals, is represented
on a grid with a radius of~$60 \bohr$ using 3001~finite-element functions.
From its eigenfunctions we choose, for each orbital angular momentum~$l$,
the 100~functions which are lowest in energy to form atomic
orbitals~\cite{Buth:TX-07}.
In doing so, we consider spherical harmonics with up
to~$l=7$~\cite{Merzbacher:QM-98,Arfken:MM-05}.
Continuum electrons are treated with a smooth exterior complex scaling
complex absorbing potential~\cite{Moiseyev:DU-98,Riss:TC-98,Karlsson:AR-98}
which is parametrized with the complex scaling angle~$\theta  = 0.13 \rad$,
a smoothness of the path of~$\lambda = 5 \bohr^{-1}$, and an exteriority
of~$r_0 = 10 \bohr$~\cite{Buth:TX-07}.
There is only radiative decay of singly excited states of helium with
comparatively long lifetimes to all other time scales in the problem;
therefore, we set the linewidth of a $K$~vacancy in helium to zero.
Finally, we diagonalize the involved Floquet-type matrices to obtain the cross section. Without the laser field this is done exactly; when the laser is present we use 4000~Lanczos iterations~\cite{Buth:AR-08}.

Computations with the \textsc{tdse}-projection method of Sec.~\ref{sec:tdpt}
were carried out with a one-electron \textsc{tdse} solver code which is based on the algorithms
described in Ref.~\onlinecite{Schafer:NM-08}.
The same potential used above is transferred to a radial grid with spacing of $0.2 \bohr$ and used for the TDSE-projection and fully time-dependent computations (see below). The interpolation of the Hartree-Fock potential onto the coarse grid used for the TDSE propagation introduces a small error in the helium 1s ionization potential which we correct by slightly changing the potential at the first grid point~\cite{Schafer:NM-08}.
The pulse shape is given by Eq.~(\ref{eq:xuvpulse}-\ref{eq:PulseDuration})
with~$n = 6$. This means that the total propagation time is 4.67$\tau\I{X}$.
Typically we use a box of 200 au in size, with a 50 au absorbing boundary at the outer edge~\onlinecite{Schafer:NM-08}.
The maximum angular momentum and time step size are adjusted to achieve convergence. We use $\ell_{\mathrm max}=8$ and 1500 steps per dressing laser cycle. In some cases where the \XUV{}~pulse was very long or the \XUV\ wavelength was very close to the ionization threshold, the box size was increased to 1000 au ensure that no wave function amplitude that might reflect from the absorbing boundary could interfere with amplitude excited at a later time.
We specify the bandwidth~$\Delta \omega\I{X}$ of the \XUV{}~pulse
instead of the \textsc{fwhm} duration~$\tau\I{X}$ as in Eq.~(\ref{eq:GaussEnv}).
For our (approximately) Gaussian pulse, we use the time-bandwidth
product~$\tau\I{X} \, \Delta \omega\I{X} = 4 \ln 2$
to convert between the two quantities~\cite{Diels:UL-06} with~$\Delta \omega\I{X} = 0.05 \eV$ which corresponds
to a duration
%
%
of~$\tau\I{X} = 36.5 \U{fs}$.
We investigated the dependence of the absorption cross sections
on the intensity of the \XUV{}~light;
to a very good approximation, we find a linear relationship as should hold
for a one-photon absorption process~(\ref{eq:xsect}).
An \XUV{}~intensity of~$10^{10} \U{\frac{W}{cm^2}}$ is employed in Figs.1-4.

The calculations with the time-frequency method of Sec.~\ref{sec:timefreq} were performed with the TDSE solver described above.  For the calculations in Figs.~1(b) and (c) and the inset in Fig.~2 we have used $\ell_{\mathrm max}=8$ and approximately 4000 steps per cycle of the dressing laser field. The size of the radial grid was 150$a_0$ (using 750 points) with a 250 point absorbing boundary. The intensity envelope of the \IR\ pulse is $\cos^4(\beta t/\tau_{IR})$, where $\tau_{IR}$ is the \textsc{fwhm} duration of the \IR\ pulse and $\beta=2\arccos(0.5^{1/4})$. The intensity envelope of the \XUV\ is usually chosen to be the fourth power of the \IR\ envelope (to be consistent with the \XUV\ being a high order harmonic produced by the \IR\ pulse). This gives a \textsc{fwhm} pulse duration for the \XUV\ pulse of about half that of the \IR\ pulse. The window function discussed in Sec.~\ref{sec:timefreq} is a Hann window with a \textsc{fwhm} duration very close to that of the \IR\ pulse. The window function was chosen such that the long-pulse calculation in Fig.~1(b) can be compared to those in Fig.~1(a): the \textsc{fwhm} bandwidth of the windowed acceleration spectrum $\tilde a_W(\omega)$ has the same 0.05~eV bandwidth as the TDSE-projection approach.

For the \textsc{mwe-tdse} calculations in Fig.~6 we employ two time scales. One time scale defines the spectral resolution of the macroscopic, propagating, electric fields. This time scale typically extends to $\pm 3$ times the \textsc{fwhm} of the longest of the \IR\ and \XUV\ pulses and contains approximately 5500 time points. The other time scale is used for the \textsc{tdse} solution and extends only over the finite duration of the longest pulse, and typically contains 6000 points per \IR\ laser cycle. The macroscopic length scales cover 160~$\mu$m in the radial direction, with 200 grid points, 40 of which contain an absorber that prevents reflections from the edge of the grid, and 1~mm in the propagation direction, with 600 grid points. In the propagation direction we only evaluate the dipole moment every 20 steps, and rescale the response to the appropriate density and phase in between, see \cite{Gaarde:MA-08} for details. The initial spatial distribution of both the \XUV\ and the dressing laser beam is Gaussian. The \XUV\ beam has  a confocal parameter of 10~cm and a corresponding focal diameter of 60~$\mu$m. The 1600~nm dressing pulse has a confocal parameter of 2~cm and a focal diameter of 140~$\mu$m.
This means that in the spatial dimension, the \XUV\ beam is always overlapped with the \IR\ beam. The \IR\ beam changes only marginally during the propagation in the helium gas.

\section{Results and discussion}
\label{sec:results}

\subsection{Single atom absorption cross sections}

\begin{figure}
 \begin{center}
   \includegraphics[clip,width=\hsize]{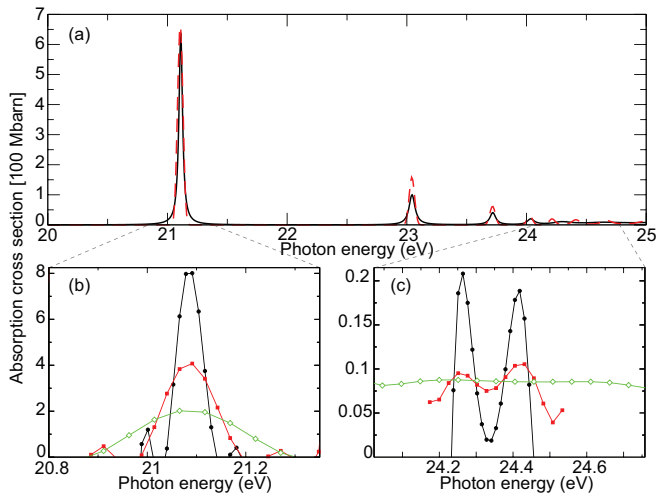}
   \caption{(Color online) The \XUV{}~absorption cross section of a helium
            atom.
            (a) The dashed red lines were obtained with  \textsc{dreyd}~\cite{fella:pgm-V1.3.0}  the solid black lines were obtained using the \textsc{tdse}-projection method.
            (b) and (c) show close-ups of the cross section around the $2p$ (b), $6p$, and $7p$ (c) states calculated
            with the time-frequency method [Eq.~(\ref{sigmaTD}] for different \XUV\ pulse durations. The results obtained using 30~fs, 15~fs, and 7.5~fs \XUV pulses are shown in black (circles), red (squares), and green (open diamonds), respectively.
           }
   \label{fig:helium_XUV}
 \end{center}
\end{figure}

The helium absorption cross section for linearly polarized \XUV{}~light, in the absence of  laser light, is displayed in Fig.~\ref{fig:helium_XUV}. In part (a), we compare results from  \textsc{dreyd}~\cite{fella:pgm-V1.3.0} with results of the  \textsc{tdse}-projection method [Eq.~(\ref{eq:excross})]. To be able to compare these two results, we have convoluted the \textsc{dreyd}~cross sections with a Gaussian with the same bandwidth of~$\Delta \omega\I{X} = 0.05 \eV$ that was used in the  TDSE calculation. This leads to good agreement between the two results. We note that the presence of a spectral bandwidth in both calculations means that we are only able to resolve spectral features to within 0.05 eV.
The peaks at~$21.1068 \eV$, $23.0416 \eV$, $23.7162 \eV$, $24.0273 \eV$, \ldots\ stem
from $1s^2 \to 1s np$~transitions with~$n \in \{2, 3, 4, 5, \ldots\}$.
In parts (b) and (c) we show cross sections calculated using the time-frequency approach  leading to Eq.~(\ref{sigmaTD}), around the $2p$ and the $6p$ and $7p$ states. These calculations were done using an extremely weak 764~nm \IR\ pulse and harmonics 13 (b) or  15 (c) of the \IR\ frequency \footnote{The \IR\ field is included as a technical convenience for performing the low-\IR\ intensity and high-\IR\ intensity calculations consistently, both for the time-frequency approach and for the \textsc{mwe-tdse} approach.}. Harmonic 13 is resonant with the $2p$ state and harmonic 15 is in between the $6p$ and the $7p$ states. We show the results of using three different \XUV\ pulse durations (30~fs, 15~fs, and 7.5~fs). The \IR\ pulse has twice the duration of the \XUV\ pulse and an intensity of $10^{8} \U{\frac{W}{cm^2}}$ (low enough that it does not influence the cross sections). The 30~fs calculation leads to a 0.05~eV bandwidth of the dipole moment around the $2p$ state, after applying the time-domain window function discussed in Sec.~\ref{sec:compdet}. The calculated cross section is in reasonably good agreement with the results in (a). The shorter \XUV\ pulses lead to broader absorption cross sections. For the 15~fs \XUV\ pulse the $6p$ and $7p$ states can still be distinguished as separate features in the absorption spectrum. Using 7.5~fs \XUV\ pulse the cross section can be calculated over a much larger frequency range, spanning both below and above the ionization threshold, and as a consequence one can no longer distinguish the $6p$ and $7p$ states. The value of the cross section in this calculation is in good agreement with the value in Henke {\it et al.} \cite{Hen93181} of 7.5~Mbarn just above threshold, as we expect when using pulses that span the ionization threshold.

\begin{figure}
 \begin{center}
   \includegraphics[clip,width=\hsize]{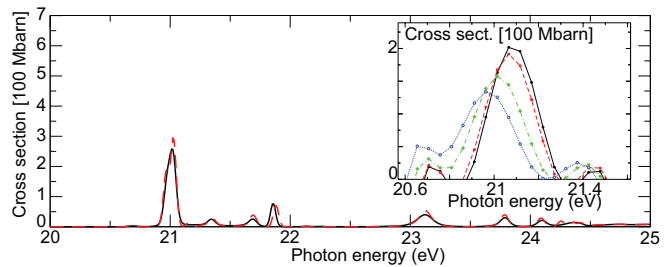}
   \caption{(Color online) The \XUV{}~absorption cross section of a helium
            atom dressed by an intense 800~nm \IR\ laser pulse with a peak intensity
            of~$10^{12} \U{\frac{W}{cm^2}}$. The dashed red lines were obtained with
            \textsc{dreyd}~\cite{fella:pgm-V1.3.0};
            the cross section obtained from the \textsc{tdse}-projection is plotted with
            dashed black lines.
            The inset shows the cross section calculated using Eq.~(\ref{sigmaTD}),
            using an \XUV\ pulse duration of 7.5~fs (\XUV\ pulse duration 7.5~fs), and an 764~nm \IR\ pulse with a duration of 15~fs and a peak
            intensity of $10^{8} \U{\frac{W}{cm^2}}$ (solid black curve), $10^{11} \U{\frac{W}{cm^2}}$ (dashed red), $5 \times 10^{11} \U{\frac{W}{cm^2}}$ (dot-dashed green), or
            $10^{12} \U{\frac{W}{cm^2}}$ (dotted blue), respectively.}
   \label{fig:helium_XUV_IR}
 \end{center}
\end{figure}

In Fig.~\ref{fig:helium_XUV_IR} we show how the \XUV\ cross section changes when the helium atom is exposed to an infrared laser field with an intensity of  $10^{12} \U{\frac{W}{cm^2}}$ and a wavelength of $\sim 800$ nm. The main figure again compares the results from  \textsc{dreyd}~\cite{fella:pgm-V1.3.0} and the \textsc{tdse}-projection method (which have again both been calculated/convoluted with a 0.05~eV bandwidth), and they are found to be in good agreement over a broad energy range.

The absorption of the dressed atom in Fig.~\ref{fig:helium_XUV_IR} changes significantly from the undressed case, although many of the field-free resonances can still be recognized. The peak due to the $2p$ state has broadened and shifted to lower energy, whereas the higher $np$ peaks are shifted to higher energies. In addition, several new absorbing features have appeared between 21~eV and 22~eV. The inset in Fig.~\ref{fig:helium_XUV_IR} shows cross sections for the dressed helium atom calculated using the time-frequency approach of Eq.~(\ref{sigmaTD}), for an \XUV\ pulse duration of 7.5~fs. The 764~nm \IR\ pulse duration is 15~fs and the \IR\ peak intensity varies between $10^{8} \U{\frac{W}{cm^2}}$ (undressed, as shown in Fig.~\ref{fig:helium_XUV}(b)) and $10^{12} \U{\frac{W}{cm^2}}$. The inset details the shift and broadening of the $2p$ resonance as the dressing laser intensity is increased. We have chosen to use the 7.5~fs \XUV\ pulses for these calculations in order to be able to cover the shift of the $2p$ resonance within the bandwidth that can be addressed within Eq.~(\ref{sigmaTD}).

The extra peaks in the 800~nm dressed-atom cross section shown in Fig.~\ref{fig:helium_XUV_IR} result from complex multiphoton effects and do not have a straight forward interpretation. Other dressing-laser wavelengths offer more insight into the non-linear optics driven by the two-color field. We first show two figures exploring the impact of the dressing laser wavelength on the \XUV{}~absorption cross section. The wavelengths we have used are listed in Table~\ref{tab:phenergy} together with the corresponding photon energies. All of these wavelengths can be produced from standard Ti:Sapphire high power, short pulse laser systems via frequency mixing in nonlinear materials.

\begin{table}
 \centering
 \begin{ruledtabular}
   \begin{tabular}{r|ccccccc}
     Wavelength [nm]    \quad &  400 &  500 &  620 &  800 & 1400 & 1600 &
                                2000 \\
     \hline
     Photon energy [eV] \quad & 3.10 & 2.48 & 2.00 & 1.55 & 0.89 & 0.78 &
                                0.62
   \end{tabular}
 \end{ruledtabular}
 \caption{Correspondance between wavelenght and photon energy for the
          involved laser light.}
 \label{tab:phenergy}
\end{table}

\begin{figure}
 \begin{center}
   \includegraphics[clip,width=\hsize]{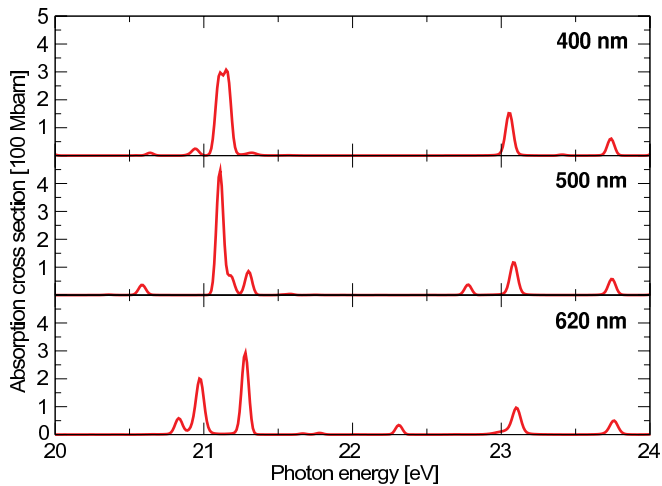}
   \caption{(Color online) Laser-dressed \XUV{}~absorption cross section
            of helium for 400, 500, and $620 \U{nm}$~laser wavelengths
            at a laser intensity of~$10^{12} \U{\frac{W}{cm^2}}$.}
   \label{fig:column_first}
 \end{center}
\end{figure}

\begin{figure}
 \begin{center}
   \includegraphics[clip,width=\hsize]{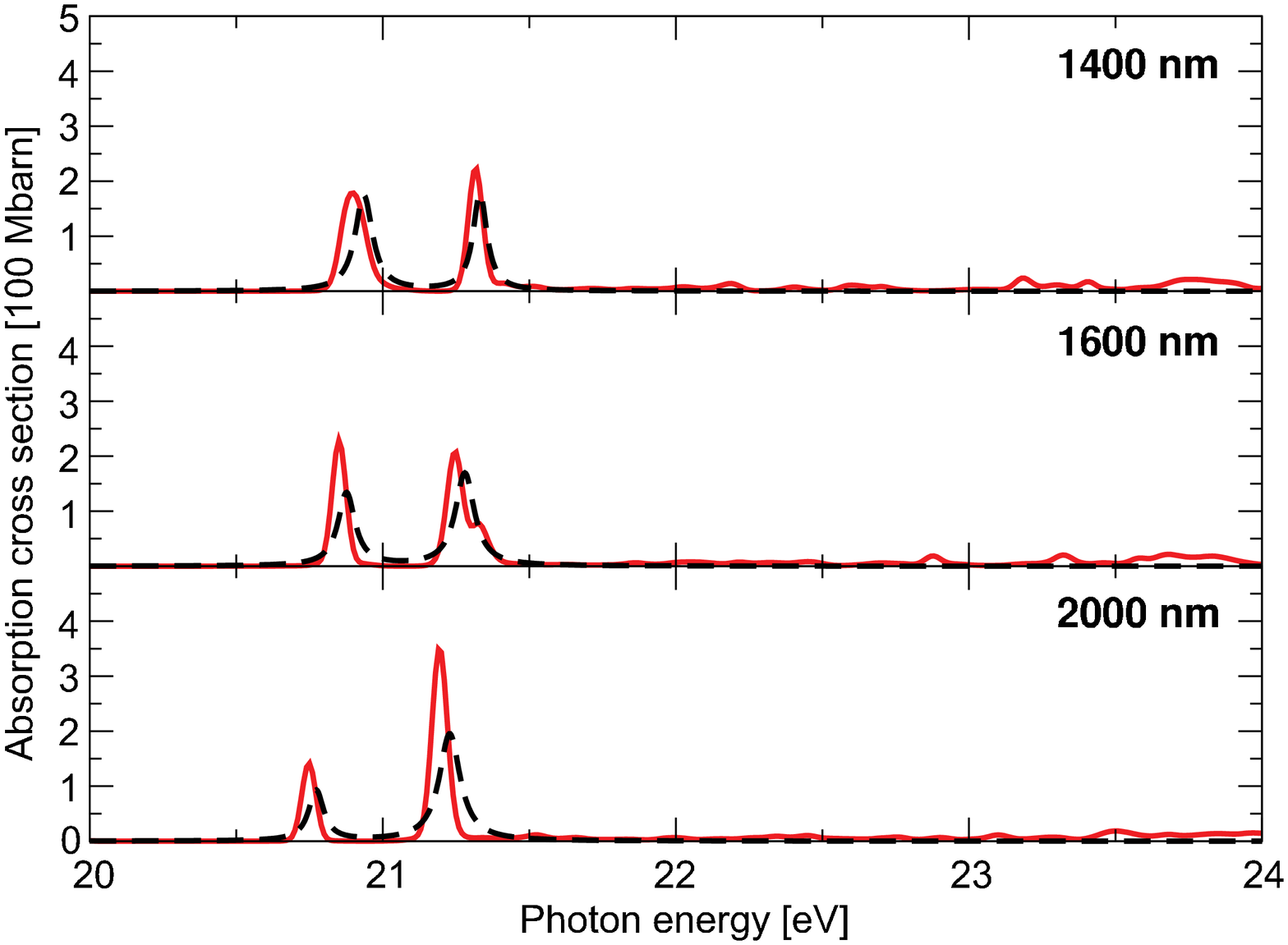}
   \caption{(Color online) Laser-dressed \XUV{}~absorption cross section
            of helium for 1400, 1600, and $2000 \U{nm}$~laser wavelengths
            at a laser intensity of~$10^{12} \U{\frac{W}{cm^2}}$
            (red solid curves).
            The results of the simple three-level model of
            Fig.~\ref{fig:three-level} are indicated
            by the dashed black curves for~$\Gamma_{1s^{-1} \, 2p}
            = 0.1 \eV$ and $\Gamma_{1s^{-1} \, 2s} = 0.05 \eV$.}
   \label{fig:column_second}
 \end{center}
\end{figure}

The laser-dressed \XUV{}~absorption cross section, calculated using the TDSE-projection method are displayed in Figs.~\ref{fig:helium_XUV_IR}, \ref{fig:column_first}, and
\ref{fig:column_second} for several laser wavelengths at an intensity of~$10^{12} \U{\frac{W}{cm^2}}$.
For $400 \U{nm}$ and $500 \U{nm}$~light, we see only a moderate impact of the laser dressing. The impact is mostly on the $2p$ state, as the largest dipole coupling exists to other close-by Rydberg states.  $620 \U{nm}$ and $800 \U{nm}$~light exhibit complex multiphoton effects which manifest in complicated multi-peak structures in the cross sections.
The dressing pulses with longer wavelengths all induce systematic behavior. In all three cases, the single $1s^2 \to 1s 2p$~transition in
Fig.~\ref{fig:helium_XUV}(a) (without dressing) is split into two lines by the laser in Fig.~\ref{fig:column_first};
the transitions from the $1s$~orbital into higher Rydberg orbitals are replaced by a continuous, weak absorption feature.

\begin{figure}
 \begin{center}
   \includegraphics[clip,width=2in]{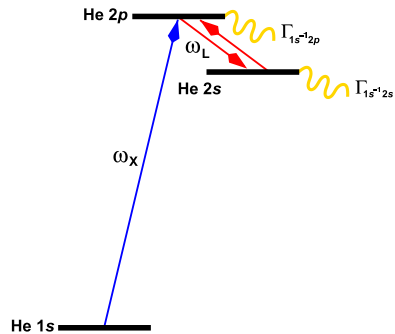}
   \caption{(Color online) $\Lambda$-type three-level model for helium.
            The laser photon energy is~$\omega\I{L}$ and
            the \XUV{}~photon energy is~$\omega\I{X}$.
            The laser-induced decay widths of the $1s^{-1} \, 2p$ and
            $1s^{-1} \, 2s$ excited states are denoted
            by~$\Gamma_{1s^{-1} \, 2p}$ and $\Gamma_{1s^{-1} \, 2s}$,
            respectively.}
   \label{fig:three-level}
 \end{center}
\end{figure}

We would like to elucidate the origin of the double-peak feature
around~$21 \eV$ in the long-wavelength series shown
in Fig.~\ref{fig:column_second}.
It is much simpler than the corresponding feature for the wavelengths
in Figs.~\ref{fig:helium_XUV_IR} and \ref{fig:column_first}.
To this end, we make a $\Lambda$-type model for helium which is shown in
Fig.~\ref{fig:three-level}.
It comprises the ground state of helium and the~$1s^{-1} \, 2p$,
$1s^{-1} \, 2s$~excited states.
The laser photon energy is denoted by~$\omega\I{L}$ whereas
$\Gamma_{1s^{-1} \, 2s}$ and $\Gamma_{1s^{-1} \, 2p}$ are parameters
for the laser-induced decay widths of the respective excited states.
The overall agreement of the model curves with the \emph{ab initio} data
in Fig.~\ref{fig:column_second} is satisfactory.
The reason for the success of the three-level model is---as in
Ref.~\onlinecite{Buth:ET-07}---the fact that the
splitting between the $2s$ and $2p$~Rydberg orbitals in helium is~$0.84 \eV$, {\it i.e.}, the laser is almost in resonance with this transition, within the laser-induced line widths, for midinfrared
wavelengths [Table~\ref{tab:phenergy}]. Furthermore, the other levels of helium couple only weakly.

The $\Lambda$-type model explains the double peaked structure in Fig.~\ref{fig:column_second}  in terms of a splitting of the~$1s^{-1} \, 2p$
and $1s^{-1} \, 2s$~states into an Autler-Townes doublet. This feature raises the possibility that the dressing laser
could be used to induce transparency to the \XUV\ radiation tuned to the $1s^2 \to 1s2p$ transition.
A similar mechanism was found for the suppression of resonant absorption
of x~rays in neon~\cite{Buth:ET-07,Santra:SF-07,Buth:RA-10,Young:US-10},
argon~\cite{Buth:AR-08}, and krypton~\cite{Buth:TX-07} atoms and called
\textsc{eit} for x~rays~\cite{Buth:ET-07}. In the next section we study the analogous effect in helium.

\subsection{\XUV\ pulse shaping in a macroscopic medium}

In this section we present an application of the time-frequency approach to absorption in a macroscopic non-linear medium. We study how  the laser induced transparency discussed above may be used to temporally control the \XUV\ pulse shape in a helium gas, in analogy with the \textsc{eit} for x-rays discussed in \cite{Buth:AR-08}. In that x-ray study, the absorption was exclusively described in terms of an intensity-dependent absorption cross section, which then in turn enforces a one-to-one mapping of the absorption to time through the intensity. This does not allow for truly dynamical effects.  In addition, the intensities we explore here are much lower that those used in the x-ray study, which means that ionization of the Rydberg states does not play a large role.

We calculate the electric field of a combined two-color \XUV{}-\IR\ pulse after propagation through a 1~mm long helium gas jet with a density of $1.5\times 10^{17}$~cm$^{-3}$ (6~mbar at room temperature). We solve the coupled \textsc{mwe-tdse} in the form of Eq.~(\ref{MWE}) as described in section \ref{Macroscopic}, see also \cite{Gaarde:MA-08}. The initial \XUV\ pulse has a wavelength of 58.7~nm (21.1~eV, resonant with the $1s^2 \to 1s 2p$ transition), a pulse duration of 61~fs and a peak intensity of $10^7$~W/cm$^2$. The 1600~nm dressing pulse has a peak intensity of $10^{12}$~W/cm$^2$. We have used different \IR\ pulse durations between 122~fs and 11~fs.  We have checked that the reshaping discussed below is no different when we use a higher \XUV\ intensity of $10^{10}$~W/cm$^2$.

\begin{figure}
   \includegraphics[width=8cm]{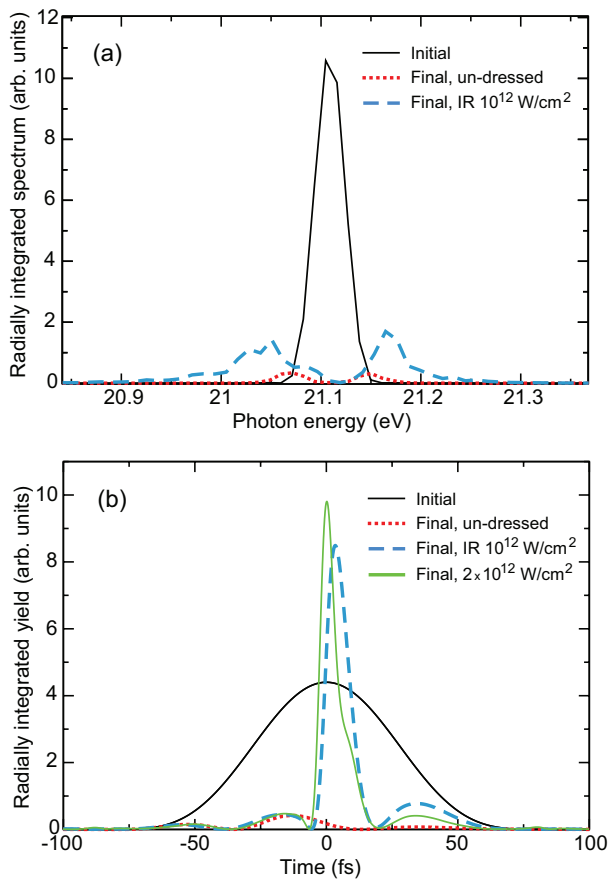}
   \caption{(Color online). \IR-assisted \XUV\ absorption in 1 mm long macroscopic helium gas with a density of $4\times 10^{16}$~cm$^{-3}$. The initial 61~fs \XUV\ pulse is resonant with the $2p$ state of the undressed helium atom. We show the \XUV\ spectrum in (a) and time profile in (b), both before (solid lines) and after propagation. Final profiles at the end of the un-dressed medium are shown with dotted red lines, final profiles at the end of the medium dressed by an 11~fs, 1600 nm \IR\ pulse with a peak intensity of $10^{12}$~W/cm$^2$ are shown with dashed blue lines. In (b) we also show the final time profile when the intensity of the dressing pulse is $2 \times 10^{12}$~W/cm$^2$ (thin green line).}
   \label{fig:absorption}
\end{figure}

Fig.~\ref{fig:absorption} shows the radially integrated spectrum (a) and time profile (b) of the \XUV\ pulse before and after propagation through the helium gas. When the atoms are undressed  the \XUV\ radiation is strongly depleted via the resonant absorption, as is shown by the dotted red lines \footnote{We note that a weak 61~fs \IR\ pulse with a peak intensity of $10^{8}$~W/cm$^2$ was present in this calculation for computational consistency with the dressed-field calculation. We have checked that the presence of this pulse does not alter the results}. The absorption length at this atomic density is less than 0.1~mm. During the first few absorption lengths the \XUV\ yield decreases exponentially. The large dispersion across the resonance, and to a lesser extent the frequency dependence of the absorption cross section, subsequently leads to reshaping of the depleted beam upon further propagation in the gas. This causes the double-peaked shape of the spectrum emerging at the end of the medium. The time profile of the final \XUV\ field is correspondingly irregular, as seen by the red dotted line in Fig.~\ref{fig:absorption}(b).

We then apply a 1600~nm, $10^{12}$~W/cm$^2$ dressing pulse which is much longer than the \XUV\ pulse (123 fs vs 61~fs). This means that the \XUV\ pulse encounters a sample of strongly dressed atoms which are no longer resonant with the \XUV\ energy, see Fig.~4, and the gas is therefore transparent to the \XUV\ light. The spectrum of the final \XUV\ pulse is nearly indistinguishable from the initial spectrum and is not shown in Fig.~\ref{fig:absorption}(a). The final \XUV\ pulse shape is also nearly identical to the initial pulse shape except for a 1.6~fs delay caused by the different group velocities of the \IR\ and the \XUV\ pulses (also not shown in the figure).

Next, we apply an 11~fs dressing \IR\ pulse which is substantially shorter than the \XUV\ pulse. This means that the dressing pulse turns on and off within the \textsc{fwhm} duration of the \XUV\ pulse, thereby strongly coupling the $2s$ and $2p$ states in a dynamical manner. The final spectral and temporal \XUV\ profiles at the end of the medium are shown with dashed blue lines in Fig.~\ref{fig:absorption}(a) and (b).  The \XUV\ time profile at the end of the medium is dominated by an approximately 10~fs pulse, superimposed on a much weaker longer pulse, and the corresponding \XUV\ spectrum has broad shoulders at frequencies substantially beyond the initial \XUV\ bandwidth. We note that the final \XUV\ pulse is not symmetric around time zero and that in particular, the short sub-pulse is delayed by approximately 4~fs from the center of the dressing  \IR~pulse. We attribute this to the complicated absorption and emission dynamics driven by the two-color pulse as explored in more detail next.

\begin{figure}
   \includegraphics[width=7cm]{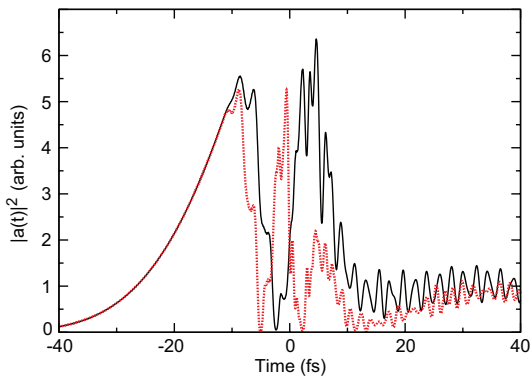}
   \caption{(Color online). Single atom time-dependent acceleration driven by the initial \XUV-\IR\ pulse in Fig.~\ref{fig:absorption}. The duration of the two pulses are 61~fs and 11~fs, respectively, and the \XUV\ intensity is $10^{7}$~W/cm$^2$. In solid black line we show the result of using an \IR\ intensity of $1 \times 10^{12}$~W/cm$^2$, in dashed red line the \IR\ intensity is $3 \times 10^{12}$~W/cm$^2$.}
   \label{fig:acc}
\end{figure}

The time-dependent acceleration driven by the initial two-color pulse is shown in Fig.~\ref{fig:acc}, solid black line. We are showing the envelope of the acceleration to avoid the fast oscillations at the resonance frequency.
On the rising edge of the \XUV\ pulse, before the \IR\ pulse turns on, this acceleration represents the absorption of the \XUV\ light via population transfer to the 2p state. This means that when the \IR\ pulse arrives there is already population in the 2p state which will couple strongly to the 2s state. This causes the suppression of the acceleration which starts around $t=-5$~fs. The acceleration then has a revival around 4~fs before it is suppressed again. By calculating the generalized cross sections of Eq.~(\ref{sigmaTD}) separately for the peaks around $t=-9$~fs and around $t=+4$~fs, we find that they have opposite signs. This means that whereas the dipole response early in the pulse and up until $t\approx -5$~fs is due to the absorption of the \XUV\ light,  the peak in Fig.~\ref{fig:acc} around  $t=+4$~fs corresponds to {\it emission} of \XUV\ radiation. We interpret this behavior as coming from Rabi-like oscillations of the excited state population between the 2s and the 2p states, driven by the \IR\ field.
The emission happens when the population returns to the 2p state while the strong \IR\ field is still on. This interpretation would predict that at higher \IR\ intensity the Rabi cycling should be faster. We indeed find that if we increase the \IR\ intensity to $3 \times 10^{12}$~W/cm$^2$, the time-dependent acceleration has two revivals within the \IR\ pulse duration, both corresponding to emission (dashed red line in Fig.~\ref{fig:acc}). The Rabi oscillation period for resonant population transfer between the 2s and the 2p states is approximately 10~fs (6~fs) for a constant intensity of $1 \times 10^{12}$~W/cm$^2$ ($3 \times 10^{12}$~W/cm$^2$). This is in good agreement with the time scale of the oscillations in the acceleration seen in Fig.~7, especially considering that the \IR\ intensity is changing rapidly between $t=-10$~fs and $t=10$~fs.

\begin{figure}
   \includegraphics[width=7cm]{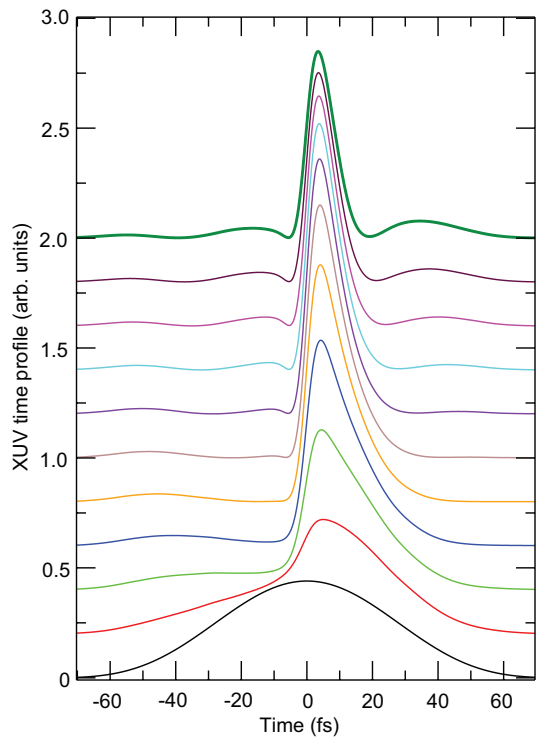}
   \caption{(Color online). Evolution of \XUV\ time profile during propagation through the macroscopic helium gas. The time profiles at different propagation distances have been displaced vertically, starting with $z=0$ at the bottom to $z=1.0$~mm at the top, in increments of $\Delta z = 0.01$~mm. }
   \label{fig:PropagEvolution}
\end{figure}

Finally, returning to Fig.~\ref{fig:absorption}(b) and the \XUV\ pulse that emerges from the helium gas dressed by the 11~fs \IR\ pulse, we can now attribute the delay of the short \XUV\ pulse to the excited state dynamics in the strongly dressed atomic gas. Fig.~\ref{fig:PropagEvolution} shows the evolution of the \XUV\ time profile shown in Fig.~\ref{fig:absorption}, as a function of propagation distance. The complicated atomic response shown in Fig.~\ref{fig:acc}, which includes absorption at early times and emission around $t=4$~fs, is reflected in the propagating \XUV\ electric field. After the first few hundred microns of propagation the \XUV\ time profile has been substantially depleted on the rising edge and is dominated by a much shorter pulse peaking shortly after $t=0$. We note that this in turn will change the atomic response from that plotted in Fig.~\ref{fig:acc} since the dressing \IR\ pulse and the \XUV\ pulse are then more comparable in duration.

\section{Conclusion}
\label{sec:conclusion}

In this paper, we have investigated the response of laser-dressed helium atoms to \XUV\ radiation,
within the \textsc{sae} approximation. In particular, we have focused on the calculation of absorption cross sections and their application to absorption in a macroscopic medium.

First, we introduced a time-independent method based on \textsc{nhpt}.
The interaction with the \XUV{}~light was treated in terms of a
one-photon process while a Floquet-like approximation was used to describe
the impact of the dressing laser.
Second, we devised a time-dependent method to compute the cross section
with  using a direct integration of
the \textsc{tdse} and projection of the final wave packet onto the initial atomic state.
We showed that the projection-based approach, which was implemented using finite pulses, yields the same results
compared with the time-independent results.
Third, we presented a versatile time-frequency approach to evaluating  an atomic response function which can be used even when the dressing laser pulse is so short that it introduces transient effects, or in cases where the atom exchanges energy with multiple frequency components of the multi-color light field. We showed that this method, when used to calculate linear absorption cross sections, agrees with the first two. Finally, we showed that this third approach can be implemented in a combined \textsc{mwe-tdse} solver to describe absorption and ultrafast pulse reshaping in a macroscopic medium.

We used the TSDE-projection method to investigate the
dependence of the \XUV{}~absorption cross section on the
wavelength of the laser dressing at
$10^{12} \U{\frac{W}{cm^2}}$~laser intensity.
We found complex multiphoton physics for $800 \U{nm}$~light and shorter
wavelengths.
For longer, midinfrared wavelengths, however, we showed that the impact
of the laser dressing in the $1s \to 2p$~transition in helium can be
described in terms of a $\Lambda$-type three-level model previously used to describe
\textsc{eit} for x-rays~\cite{Buth:ET-07}.
As in the earlier study,
the transparency in helium is caused predominantly by Autler-Townes splitting brought about by the strong one photon coupling induced by the dressing laser, in this case between the $2p$ and $2s$ states.
We investigated the macroscopic reshaping of an ultrafast \XUV\ pulse resonant with this transition, for the case when the transparency is induced by an \IR\ pulse which is substantially shorter than the \XUV\ pulse. This means that the absorption properties of the helium atom change dynamically during its interaction with the \XUV\ light. We found rich temporal reshaping dynamics in which the atoms both absorb and subsequently emit the \XUV\ radiation in a process strongly influenced by Rabi oscillations between the $2s$ and $2p$ states. This leads to an \XUV\ pulse emerging from the macroscopic medium which has been shortened from 60~fs to 10~fs and whose peak intensity has increased by approximately a factor of two. The increase in the peak intensity, which results from the coherent population pumped into the $2p$ state before the dressing pulse arrives, is a truly dynamical effect which cannot be described in terms of a single absorption cross section only.

Our results open up several possibilities for future research on ultrafast quantum optics.
The  control of \XUV{}~absorption by laser dressing
of helium enables, for example, the possibility for post-generation ultrafast shaping of
\XUV{}~pulses~\cite{Buth:ET-07}. And although we have confined ourselves in this work to the case
of one \XUV\ field with a dressing laser, it is a straightforward extension of the method to treat multiple \XUV\ frequencies some of which some could be resonant with dressed atomic transitions, and the complex interferences that would result from this \cite{Johnsson:AC-07, Ranitovic2010013008}.
Also, we have used moderately strong IR fields that do not cause any excitation on their own, but the time-dependent treatment is not limited to these intensities. Using higher IR intensities will lead to generation of harmonics in the non-linear medium. Harmonics with energies below and slightly above the ionization threshold, for which absorption dynamics plays the largest role, have recently attracted a lot of attention, for instance as a source of \textsc{vuv} and \XUV\ frequency combs \cite{Yost:BT-09}  or as a seed for free electron lasers \cite{He09063829}. The treatment of these processes necessitates using methods we have developed in this paper, since the atomic absorption and emission properties will be changing on an ultrafast time scale.

\begin{acknowledgments}
This work was supported by the National Science Foundation under
grant~Nos.~PHY-0449235 (CB and JT), PHY-0701372 (KS), and PHY-1019071 (MG). We also acknowledge support from the PULSE Institute at Stanford University (MG and KS) and from the  Marie Curie International Reintegration Grant  (CB) within the 7$^{\mathrm{th}}$~European Community Framework Program (call identifier: FP7-PEOPLE-2010-RG, proposal No.~266551). High performance computational resources were provided by the Louisiana Optical
Network Initiative, www.loni.org.
\end{acknowledgments}


\begin{thebibliography}{52}
\expandafter\ifx\csname natexlab\endcsname\relax\def\natexlab#1{#1}\fi
\expandafter\ifx\csname bibnamefont\endcsname\relax
  \def\bibnamefont#1{#1}\fi
\expandafter\ifx\csname bibfnamefont\endcsname\relax
  \def\bibfnamefont#1{#1}\fi
\expandafter\ifx\csname citenamefont\endcsname\relax
  \def\citenamefont#1{#1}\fi
\expandafter\ifx\csname url\endcsname\relax
  \def\url#1{\texttt{#1}}\fi
\expandafter\ifx\csname urlprefix\endcsname\relax\def\urlprefix{URL }\fi
\providecommand{\bibinfo}[2]{#2}
\providecommand{\eprint}[2][]{\url{#2}}

\bibitem[{\citenamefont{Johnsson et~al.}(2007)\citenamefont{Johnsson,
  Mauritsson, Remetter, L'Huillier, and Schafer}}]{Johnsson:AC-07}
\bibinfo{author}{\bibfnamefont{P.}~\bibnamefont{Johnsson}},
  \bibinfo{author}{\bibfnamefont{J.}~\bibnamefont{Mauritsson}},
  \bibinfo{author}{\bibfnamefont{T.}~\bibnamefont{Remetter}},
  \bibinfo{author}{\bibfnamefont{A.}~\bibnamefont{L'Huillier}},
  \bibnamefont{and} \bibinfo{author}{\bibfnamefont{K.~J.}
  \bibnamefont{Schafer}}, \bibinfo{journal}{Phys. Rev. Lett.}
  \textbf{\bibinfo{volume}{99}}, \bibinfo{pages}{233001}
  (\bibinfo{year}{2007}).

\bibitem[{\citenamefont{Glover et~al.}(2009)\citenamefont{Glover, Hertlein,
  Southworth, Allison, van Tilborg, Kanter, Kr\"assig, Varma, Rude, Santra
  et~al.}}]{Glover:CX-09}
\bibinfo{author}{\bibfnamefont{T.~E.} \bibnamefont{Glover}},
  \bibinfo{author}{\bibfnamefont{M.~P.} \bibnamefont{Hertlein}},
  \bibinfo{author}{\bibfnamefont{S.~H.} \bibnamefont{Southworth}},
  \bibinfo{author}{\bibfnamefont{T.~K.} \bibnamefont{Allison}},
  \bibinfo{author}{\bibfnamefont{J.}~\bibnamefont{van Tilborg}},
  \bibinfo{author}{\bibfnamefont{E.~P.} \bibnamefont{Kanter}},
  \bibinfo{author}{\bibfnamefont{B.}~\bibnamefont{Kr\"assig}},
  \bibinfo{author}{\bibfnamefont{H.~R.} \bibnamefont{Varma}},
  \bibinfo{author}{\bibfnamefont{B.}~\bibnamefont{Rude}},
  \bibinfo{author}{\bibfnamefont{R.}~\bibnamefont{Santra}},
  \bibnamefont{et~al.}, \bibinfo{journal}{Nature Phys.}
  \textbf{\bibinfo{volume}{6}}, \bibinfo{pages}{69} (\bibinfo{year}{2009}).

\bibitem[{\citenamefont{Ranitovic et~al.}(2010)\citenamefont{Ranitovic, Tong,
  Gramkow, De, DePaola, Singh, Cao, Magrakvelidze, Ray, Bocharova
  et~al.}}]{Ranitovic2010013008}
\bibinfo{author}{\bibfnamefont{P.}~\bibnamefont{Ranitovic}},
  \bibinfo{author}{\bibfnamefont{X.~M.} \bibnamefont{Tong}},
  \bibinfo{author}{\bibfnamefont{B.}~\bibnamefont{Gramkow}},
  \bibinfo{author}{\bibfnamefont{S.}~\bibnamefont{De}},
  \bibinfo{author}{\bibfnamefont{B.}~\bibnamefont{DePaola}},
  \bibinfo{author}{\bibfnamefont{K.~P.} \bibnamefont{Singh}},
  \bibinfo{author}{\bibfnamefont{W.}~\bibnamefont{Cao}},
  \bibinfo{author}{\bibfnamefont{M.}~\bibnamefont{Magrakvelidze}},
  \bibinfo{author}{\bibfnamefont{D.}~\bibnamefont{Ray}},
  \bibinfo{author}{\bibfnamefont{I.}~\bibnamefont{Bocharova}},
  \bibnamefont{et~al.}, \bibinfo{journal}{New Journal of Physics}
  \textbf{\bibinfo{volume}{12}}, \bibinfo{pages}{013008}
  (\bibinfo{year}{2010}).

\bibitem[{\citenamefont{Goulielmakis et~al.}(2010)\citenamefont{Goulielmakis,
  Loh, Wirth, Santra, Rohringer, Yakovlev, Zherebtsov, Pfeifer, Azzeer, Kling
  et~al.}}]{Goulielmakis:RO-10}
\bibinfo{author}{\bibfnamefont{E.}~\bibnamefont{Goulielmakis}},
  \bibinfo{author}{\bibfnamefont{Z.-H.} \bibnamefont{Loh}},
  \bibinfo{author}{\bibfnamefont{A.}~\bibnamefont{Wirth}},
  \bibinfo{author}{\bibfnamefont{R.}~\bibnamefont{Santra}},
  \bibinfo{author}{\bibfnamefont{N.}~\bibnamefont{Rohringer}},
  \bibinfo{author}{\bibfnamefont{V.~S.} \bibnamefont{Yakovlev}},
  \bibinfo{author}{\bibfnamefont{S.}~\bibnamefont{Zherebtsov}},
  \bibinfo{author}{\bibfnamefont{T.}~\bibnamefont{Pfeifer}},
  \bibinfo{author}{\bibfnamefont{A.~M.} \bibnamefont{Azzeer}},
  \bibinfo{author}{\bibfnamefont{M.~F.} \bibnamefont{Kling}},
  \bibnamefont{et~al.}, \bibinfo{journal}{Nature}
  \textbf{\bibinfo{volume}{466}}, \bibinfo{pages}{739} (\bibinfo{year}{2010}).

\bibitem[{\citenamefont{J. et~al.}(2010)\citenamefont{J., T., M., Kluender,
  L'Huillier, Schafer, Ghafur, Kelkensberg, Siu, Johnsson
  et~al.}}]{Mauritsson2010053001}
\bibinfo{author}{\bibfnamefont{M.}~\bibnamefont{J.}},
  \bibinfo{author}{\bibfnamefont{R.}~\bibnamefont{T.}},
  \bibinfo{author}{\bibfnamefont{S.}~\bibnamefont{M.}},
  \bibinfo{author}{\bibfnamefont{K.}~\bibnamefont{Kluender}},
  \bibinfo{author}{\bibfnamefont{A.}~\bibnamefont{L'Huillier}},
  \bibinfo{author}{\bibfnamefont{K.~J.} \bibnamefont{Schafer}},
  \bibinfo{author}{\bibfnamefont{O.}~\bibnamefont{Ghafur}},
  \bibinfo{author}{\bibfnamefont{F.}~\bibnamefont{Kelkensberg}},
  \bibinfo{author}{\bibfnamefont{W.}~\bibnamefont{Siu}},
  \bibinfo{author}{\bibfnamefont{P.}~\bibnamefont{Johnsson}},
  \bibnamefont{et~al.}, \bibinfo{journal}{Physical Review Letters}
  \textbf{\bibinfo{volume}{105}}, \bibinfo{pages}{053001}
  (\bibinfo{year}{2010}).

\bibitem[{\citenamefont{Wickenhauser et~al.}(2005)\citenamefont{Wickenhauser,
  Burgd\"orfer, Krausz, and Drescher}}]{Wickenhauser:TR-05}
\bibinfo{author}{\bibfnamefont{M.}~\bibnamefont{Wickenhauser}},
  \bibinfo{author}{\bibfnamefont{J.}~\bibnamefont{Burgd\"orfer}},
  \bibinfo{author}{\bibfnamefont{F.}~\bibnamefont{Krausz}}, \bibnamefont{and}
  \bibinfo{author}{\bibfnamefont{M.}~\bibnamefont{Drescher}},
  \bibinfo{journal}{Phys. Rev. Lett.} \textbf{\bibinfo{volume}{94}},
  \bibinfo{pages}{023002} (\bibinfo{year}{2005}).

\bibitem[{\citenamefont{Buth et~al.}(2007)\citenamefont{Buth, Santra, and
  Young}}]{Buth:ET-07}
\bibinfo{author}{\bibfnamefont{C.}~\bibnamefont{Buth}},
  \bibinfo{author}{\bibfnamefont{R.}~\bibnamefont{Santra}}, \bibnamefont{and}
  \bibinfo{author}{\bibfnamefont{L.}~\bibnamefont{Young}},
  \bibinfo{journal}{Phys. Rev. Lett.} \textbf{\bibinfo{volume}{98}},
  \bibinfo{pages}{253001} (\bibinfo{year}{2007}),
  \bibinfo{note}{\href{http://arxiv.org/abs/0705.3615} {arXiv:0705.3615}}.

\bibitem[{\citenamefont{Pfeifer et~al.}(2008)\citenamefont{Pfeifer, Abel,
  Nagel, Jullien, Loh, Bell, Neumark, and Leone}}]{Pfeifer200811}
\bibinfo{author}{\bibfnamefont{T.}~\bibnamefont{Pfeifer}},
  \bibinfo{author}{\bibfnamefont{M.~J.} \bibnamefont{Abel}},
  \bibinfo{author}{\bibfnamefont{P.~M.} \bibnamefont{Nagel}},
  \bibinfo{author}{\bibfnamefont{A.}~\bibnamefont{Jullien}},
  \bibinfo{author}{\bibfnamefont{Z.-H.} \bibnamefont{Loh}},
  \bibinfo{author}{\bibfnamefont{M.~J.} \bibnamefont{Bell}},
  \bibinfo{author}{\bibfnamefont{D.~M.} \bibnamefont{Neumark}},
  \bibnamefont{and} \bibinfo{author}{\bibfnamefont{S.~R.} \bibnamefont{Leone}},
  \bibinfo{journal}{Chemical Physics Letters} \textbf{\bibinfo{volume}{463}},
  \bibinfo{pages}{11} (\bibinfo{year}{2008}).

\bibitem[{\citenamefont{Drescher et~al.}(2002)\citenamefont{Drescher,
  Hentschel, Kienberger, Uiberacker, Yakovlev, Scrinzi, Westerwalbesloh,
  Kleineberg, Heinzmann, and Krausz}}]{Drescher:TR-02}
\bibinfo{author}{\bibfnamefont{M.}~\bibnamefont{Drescher}},
  \bibinfo{author}{\bibfnamefont{M.}~\bibnamefont{Hentschel}},
  \bibinfo{author}{\bibfnamefont{R.}~\bibnamefont{Kienberger}},
  \bibinfo{author}{\bibfnamefont{M.}~\bibnamefont{Uiberacker}},
  \bibinfo{author}{\bibfnamefont{V.~S.} \bibnamefont{Yakovlev}},
  \bibinfo{author}{\bibfnamefont{A.}~\bibnamefont{Scrinzi}},
  \bibinfo{author}{\bibfnamefont{T.}~\bibnamefont{Westerwalbesloh}},
  \bibinfo{author}{\bibfnamefont{U.}~\bibnamefont{Kleineberg}},
  \bibinfo{author}{\bibfnamefont{U.}~\bibnamefont{Heinzmann}},
  \bibnamefont{and} \bibinfo{author}{\bibfnamefont{F.}~\bibnamefont{Krausz}},
  \bibinfo{journal}{Nature} \textbf{\bibinfo{volume}{419}},
  \bibinfo{pages}{803} (\bibinfo{year}{2002}).

\bibitem[{\citenamefont{Remetter et~al.}(2006)\citenamefont{Remetter, Johnsson,
  Mauritsson, Varj, Ni, L\'epine, Gustafsson, Kling, Khan, L\'opez-Martens
  et~al.}}]{Remetter10323}
\bibinfo{author}{\bibfnamefont{T.}~\bibnamefont{Remetter}},
  \bibinfo{author}{\bibfnamefont{P.}~\bibnamefont{Johnsson}},
  \bibinfo{author}{\bibfnamefont{J.}~\bibnamefont{Mauritsson}},
  \bibinfo{author}{\bibfnamefont{K.}~\bibnamefont{Varj}},
  \bibinfo{author}{\bibfnamefont{Y.}~\bibnamefont{Ni}},
  \bibinfo{author}{\bibfnamefont{F.}~\bibnamefont{L\'epine}},
  \bibinfo{author}{\bibfnamefont{E.}~\bibnamefont{Gustafsson}},
  \bibinfo{author}{\bibfnamefont{M.}~\bibnamefont{Kling}},
  \bibinfo{author}{\bibfnamefont{J.}~\bibnamefont{Khan}},
  \bibinfo{author}{\bibfnamefont{R.}~\bibnamefont{L\'opez-Martens}},
  \bibnamefont{et~al.}, \bibinfo{journal}{Nature Physics}
  \textbf{\bibinfo{volume}{2}}, \bibinfo{pages}{323} (\bibinfo{year}{2006}).

\bibitem[{\citenamefont{Boller et~al.}(1991)\citenamefont{Boller, Imamo\v{g}lu,
  and Harris}}]{Boller:ET-91}
\bibinfo{author}{\bibfnamefont{K.-J.} \bibnamefont{Boller}},
  \bibinfo{author}{\bibfnamefont{A.}~\bibnamefont{Imamo\v{g}lu}},
  \bibnamefont{and} \bibinfo{author}{\bibfnamefont{S.~E.}
  \bibnamefont{Harris}}, \bibinfo{journal}{Phys. Rev. Lett.}
  \textbf{\bibinfo{volume}{66}}, \bibinfo{pages}{2593} (\bibinfo{year}{1991}).

\bibitem[{\citenamefont{Fleischhauer et~al.}(2005)\citenamefont{Fleischhauer,
  Imamo\v{g}lu, and Marangos}}]{Fleischhauer:ET-05}
\bibinfo{author}{\bibfnamefont{M.}~\bibnamefont{Fleischhauer}},
  \bibinfo{author}{\bibfnamefont{A.}~\bibnamefont{Imamo\v{g}lu}},
  \bibnamefont{and} \bibinfo{author}{\bibfnamefont{J.~P.}
  \bibnamefont{Marangos}}, \bibinfo{journal}{Rev. Mod. Phys.}
  \textbf{\bibinfo{volume}{77}}, \bibinfo{pages}{633} (\bibinfo{year}{2005}).

\bibitem[{\citenamefont{Buth and Schafer}(2009)}]{Buth:TA-09}
\bibinfo{author}{\bibfnamefont{C.}~\bibnamefont{Buth}} \bibnamefont{and}
  \bibinfo{author}{\bibfnamefont{K.~J.} \bibnamefont{Schafer}},
  \bibinfo{journal}{Phys. Rev. A} \textbf{\bibinfo{volume}{80}},
  \bibinfo{pages}{033410} (\bibinfo{year}{2009}),
  \bibinfo{note}{\href{http://arxiv.org/abs/0905.3756} {arXiv:0905.3756}}.

\bibitem[{\citenamefont{Swoboda et~al.}(2010)\citenamefont{Swoboda, Fordell,
  Kl\"under, Dahlstr\"om, Miranda, Buth, Schafer, Mauritsson, L'Huillier, and
  Gisselbrecht}}]{Swoboda:PM-10}
\bibinfo{author}{\bibfnamefont{M.}~\bibnamefont{Swoboda}},
  \bibinfo{author}{\bibfnamefont{T.}~\bibnamefont{Fordell}},
  \bibinfo{author}{\bibfnamefont{K.}~\bibnamefont{Kl\"under}},
  \bibinfo{author}{\bibfnamefont{J.~M.} \bibnamefont{Dahlstr\"om}},
  \bibinfo{author}{\bibfnamefont{M.}~\bibnamefont{Miranda}},
  \bibinfo{author}{\bibfnamefont{C.}~\bibnamefont{Buth}},
  \bibinfo{author}{\bibfnamefont{K.~J.} \bibnamefont{Schafer}},
  \bibinfo{author}{\bibfnamefont{J.}~\bibnamefont{Mauritsson}},
  \bibinfo{author}{\bibfnamefont{A.}~\bibnamefont{L'Huillier}},
  \bibnamefont{and}
  \bibinfo{author}{\bibfnamefont{M.}~\bibnamefont{Gisselbrecht}},
  \bibinfo{journal}{Phys. Rev. Lett.} \textbf{\bibinfo{volume}{104}},
  \bibinfo{pages}{103003} (\bibinfo{year}{2010}),
  \bibinfo{note}{\href{http://arxiv.org/abs/1002.2550} {arXiv:1002.2550}}.

\bibitem[{\citenamefont{Santra et~al.}(2007)\citenamefont{Santra, Buth,
  Peterson, Dunford, Kanter, Kr\"assig, Southworth, and Young}}]{Santra:SF-07}
\bibinfo{author}{\bibfnamefont{R.}~\bibnamefont{Santra}},
  \bibinfo{author}{\bibfnamefont{C.}~\bibnamefont{Buth}},
  \bibinfo{author}{\bibfnamefont{E.~R.} \bibnamefont{Peterson}},
  \bibinfo{author}{\bibfnamefont{R.~W.} \bibnamefont{Dunford}},
  \bibinfo{author}{\bibfnamefont{E.~P.} \bibnamefont{Kanter}},
  \bibinfo{author}{\bibfnamefont{B.}~\bibnamefont{Kr\"assig}},
  \bibinfo{author}{\bibfnamefont{S.~H.} \bibnamefont{Southworth}},
  \bibnamefont{and} \bibinfo{author}{\bibfnamefont{L.}~\bibnamefont{Young}},
  \bibinfo{journal}{J. Phys.: Conf. Ser.} \textbf{\bibinfo{volume}{88}},
  \bibinfo{pages}{012052} (\bibinfo{year}{2007}),
  \bibinfo{note}{\href{http://arxiv.org/abs/0712.2556} {arXiv:0712.2556}}.

\bibitem[{\citenamefont{Buth et~al.}(2010)\citenamefont{Buth, Santra, and
  Young}}]{Buth:RA-10}
\bibinfo{author}{\bibfnamefont{C.}~\bibnamefont{Buth}},
  \bibinfo{author}{\bibfnamefont{R.}~\bibnamefont{Santra}}, \bibnamefont{and}
  \bibinfo{author}{\bibfnamefont{L.}~\bibnamefont{Young}},
  \bibinfo{journal}{Rev. Mex. F\'is. S} \textbf{\bibinfo{volume}{56}},
  \bibinfo{pages}{59} (\bibinfo{year}{2010}),
  \bibinfo{note}{\href{http://arxiv.org/abs/0805.2619} {arXiv:0805.2619}}.

\bibitem[{\citenamefont{Young et~al.}(2010)\citenamefont{Young, Buth, Dunford,
  Ho, Kanter, Kr\"assig, Peterson, Rohringer, Santra, and
  Southworth}}]{Young:US-10}
\bibinfo{author}{\bibfnamefont{L.}~\bibnamefont{Young}},
  \bibinfo{author}{\bibfnamefont{C.}~\bibnamefont{Buth}},
  \bibinfo{author}{\bibfnamefont{R.~W.} \bibnamefont{Dunford}},
  \bibinfo{author}{\bibfnamefont{P.~J.} \bibnamefont{Ho}},
  \bibinfo{author}{\bibfnamefont{E.~P.} \bibnamefont{Kanter}},
  \bibinfo{author}{\bibfnamefont{B.}~\bibnamefont{Kr\"assig}},
  \bibinfo{author}{\bibfnamefont{E.~R.} \bibnamefont{Peterson}},
  \bibinfo{author}{\bibfnamefont{N.}~\bibnamefont{Rohringer}},
  \bibinfo{author}{\bibfnamefont{R.}~\bibnamefont{Santra}}, \bibnamefont{and}
  \bibinfo{author}{\bibfnamefont{S.~H.} \bibnamefont{Southworth}},
  \bibinfo{journal}{Rev. Mex. F\'is. S} \textbf{\bibinfo{volume}{56}},
  \bibinfo{pages}{11} (\bibinfo{year}{2010}),
  \bibinfo{note}{\href{http://arxiv.org/abs/0809.3537} {arXiv:0809.3537}}.

\bibitem[{\citenamefont{Buth and Santra}(2008{\natexlab{a}})}]{Buth:AR-08}
\bibinfo{author}{\bibfnamefont{C.}~\bibnamefont{Buth}} \bibnamefont{and}
  \bibinfo{author}{\bibfnamefont{R.}~\bibnamefont{Santra}},
  \bibinfo{journal}{Phys. Rev. A} \textbf{\bibinfo{volume}{78}},
  \bibinfo{pages}{043409} (\bibinfo{year}{2008}{\natexlab{a}}),
  \bibinfo{note}{\href{http://arxiv.org/abs/0809.3249} {arXiv:0809.3249}}.

\bibitem[{\citenamefont{Buth and Santra}(2007)}]{Buth:TX-07}
\bibinfo{author}{\bibfnamefont{C.}~\bibnamefont{Buth}} \bibnamefont{and}
  \bibinfo{author}{\bibfnamefont{R.}~\bibnamefont{Santra}},
  \bibinfo{journal}{Phys. Rev. A} \textbf{\bibinfo{volume}{75}},
  \bibinfo{pages}{033412} (\bibinfo{year}{2007}),
  \bibinfo{note}{\href{http://arxiv.org/abs/physics/0611122}
  {arXiv:physics/0611122}}.

\bibitem[{\citenamefont{Laarmann et~al.}(2005)\citenamefont{Laarmann,
  de~Castro, G\"urtler, Laasch, Schulz, Wabnitz, and
  M\"oller}}]{Laarmann:PH-05}
\bibinfo{author}{\bibfnamefont{T.}~\bibnamefont{Laarmann}},
  \bibinfo{author}{\bibfnamefont{A.~R.~B.} \bibnamefont{de~Castro}},
  \bibinfo{author}{\bibfnamefont{P.}~\bibnamefont{G\"urtler}},
  \bibinfo{author}{\bibfnamefont{W.}~\bibnamefont{Laasch}},
  \bibinfo{author}{\bibfnamefont{J.}~\bibnamefont{Schulz}},
  \bibinfo{author}{\bibfnamefont{H.}~\bibnamefont{Wabnitz}}, \bibnamefont{and}
  \bibinfo{author}{\bibfnamefont{T.}~\bibnamefont{M\"oller}},
  \bibinfo{journal}{Phys. Rev. A} \textbf{\bibinfo{volume}{72}},
  \bibinfo{pages}{023409} (\bibinfo{year}{2005}).

\bibitem[{\citenamefont{Charalambidis et~al.}(2006)\citenamefont{Charalambidis,
  Tzallas, Papadogiannis, Nikolopoulos, Benis, and
  Tsakiris}}]{Charalambidis:CPH-06}
\bibinfo{author}{\bibfnamefont{D.}~\bibnamefont{Charalambidis}},
  \bibinfo{author}{\bibfnamefont{P.}~\bibnamefont{Tzallas}},
  \bibinfo{author}{\bibfnamefont{N.~A.} \bibnamefont{Papadogiannis}},
  \bibinfo{author}{\bibfnamefont{L.~A.~A.} \bibnamefont{Nikolopoulos}},
  \bibinfo{author}{\bibfnamefont{E.~P.} \bibnamefont{Benis}}, \bibnamefont{and}
  \bibinfo{author}{\bibfnamefont{G.~D.} \bibnamefont{Tsakiris}},
  \bibinfo{journal}{Phys. Rev. A} \textbf{\bibinfo{volume}{74}},
  \bibinfo{pages}{037401} (\bibinfo{year}{2006}).

\bibitem[{\citenamefont{Laarmann et~al.}(2006)\citenamefont{Laarmann,
  de~Castro, G\"urtler, Laasch, Schulz, Wabnitz, and
  M\"oller}}]{Laarmann:RCPH-06}
\bibinfo{author}{\bibfnamefont{T.}~\bibnamefont{Laarmann}},
  \bibinfo{author}{\bibfnamefont{A.~R.~B.} \bibnamefont{de~Castro}},
  \bibinfo{author}{\bibfnamefont{P.}~\bibnamefont{G\"urtler}},
  \bibinfo{author}{\bibfnamefont{W.}~\bibnamefont{Laasch}},
  \bibinfo{author}{\bibfnamefont{J.}~\bibnamefont{Schulz}},
  \bibinfo{author}{\bibfnamefont{H.}~\bibnamefont{Wabnitz}}, \bibnamefont{and}
  \bibinfo{author}{\bibfnamefont{T.}~\bibnamefont{M\"oller}},
  \bibinfo{journal}{Phys. Rev. A} \textbf{\bibinfo{volume}{74}},
  \bibinfo{pages}{037402} (\bibinfo{year}{2006}).

\bibitem[{\citenamefont{Cormier and Lambropoulos}(1997)}]{Cormier:ED-97}
\bibinfo{author}{\bibfnamefont{E.}~\bibnamefont{Cormier}} \bibnamefont{and}
  \bibinfo{author}{\bibfnamefont{P.}~\bibnamefont{Lambropoulos}},
  \bibinfo{journal}{J. Phys. B} \textbf{\bibinfo{volume}{30}},
  \bibinfo{pages}{3095} (\bibinfo{year}{1997}).

\bibitem[{\citenamefont{Tannor}(2007)}]{Tannor07}
\bibinfo{author}{\bibfnamefont{D.~J.} \bibnamefont{Tannor}},
  \emph{\bibinfo{title}{Introduction to Quantum Mechanics: A time-dependent
  perspective}} (\bibinfo{publisher}{University Science Books},
  \bibinfo{address}{Sausalito, CA}, \bibinfo{year}{2007}).

\bibitem[{\citenamefont{Pollard and A.}(1992)}]{Pollard1992}
\bibinfo{author}{\bibfnamefont{W.~T.} \bibnamefont{Pollard}} \bibnamefont{and}
  \bibinfo{author}{\bibfnamefont{M.~R.} \bibnamefont{A.}},
  \bibinfo{journal}{Annual Reviews of Physical Chemistry}
  \textbf{\bibinfo{volume}{43}}, \bibinfo{pages}{497} (\bibinfo{year}{1992}).

\bibitem[{\citenamefont{Szabo and Ostlund}(1989)}]{Szabo:MQC-89}
\bibinfo{author}{\bibfnamefont{A.}~\bibnamefont{Szabo}} \bibnamefont{and}
  \bibinfo{author}{\bibfnamefont{N.~S.} \bibnamefont{Ostlund}},
  \emph{\bibinfo{title}{Modern quantum chemistry: Introduction to advanced
  electronic structure theory}} (\bibinfo{publisher}{McGraw-Hill},
  \bibinfo{address}{New York}, \bibinfo{year}{1989}), \bibinfo{edition}{{1st,
  revised}} ed., ISBN \bibinfo{isbn}{0-486-69186-1}.

\bibitem[{\citenamefont{Buth et~al.}(2004)\citenamefont{Buth, Santra, and
  Cederbaum}}]{Buth:NH-04}
\bibinfo{author}{\bibfnamefont{C.}~\bibnamefont{Buth}},
  \bibinfo{author}{\bibfnamefont{R.}~\bibnamefont{Santra}}, \bibnamefont{and}
  \bibinfo{author}{\bibfnamefont{L.~S.} \bibnamefont{Cederbaum}},
  \bibinfo{journal}{Phys. Rev.~A} \textbf{\bibinfo{volume}{69}},
  \bibinfo{pages}{032505} (\bibinfo{year}{2004}),
  \bibinfo{note}{\href{http://arxiv.org/abs/physics/0401081}
  {arXiv:physics/0401081}}.

\bibitem[{\citenamefont{Slater}(1951)}]{Slater:AS-51}
\bibinfo{author}{\bibfnamefont{J.~C.} \bibnamefont{Slater}},
  \bibinfo{journal}{Phys. Rev.} \textbf{\bibinfo{volume}{81}},
  \bibinfo{pages}{385} (\bibinfo{year}{1951}).

\bibitem[{\citenamefont{Slater and Johnson}(1972)}]{Slater:XA-72}
\bibinfo{author}{\bibfnamefont{J.~C.} \bibnamefont{Slater}} \bibnamefont{and}
  \bibinfo{author}{\bibfnamefont{K.~H.} \bibnamefont{Johnson}},
  \bibinfo{journal}{Phys. Rev.~B} \textbf{\bibinfo{volume}{5}},
  \bibinfo{pages}{844} (\bibinfo{year}{1972}).

\bibitem[{\citenamefont{Kukulin et~al.}(1989)\citenamefont{Kukulin,
  Krasnopol'sky, and Hor\'a\v{c}ek}}]{Kukulin:TR-89}
\bibinfo{author}{\bibfnamefont{V.~I.} \bibnamefont{Kukulin}},
  \bibinfo{author}{\bibfnamefont{V.~M.} \bibnamefont{Krasnopol'sky}},
  \bibnamefont{and}
  \bibinfo{author}{\bibfnamefont{J.}~\bibnamefont{Hor\'a\v{c}ek}},
  \emph{\bibinfo{title}{Theory of resonances}} (\bibinfo{publisher}{Kluwer},
  \bibinfo{address}{Dordrecht}, \bibinfo{year}{1989}), ISBN
  \bibinfo{isbn}{90-277-2364-8}.

\bibitem[{\citenamefont{Moiseyev}(1998{\natexlab{a}})}]{Moiseyev:CS-98}
\bibinfo{author}{\bibfnamefont{N.}~\bibnamefont{Moiseyev}},
  \bibinfo{journal}{Phys. Rep.} \textbf{\bibinfo{volume}{302}},
  \bibinfo{pages}{211} (\bibinfo{year}{1998}{\natexlab{a}}).

\bibitem[{\citenamefont{Santra and Cederbaum}(2002)}]{Santra:NH-02}
\bibinfo{author}{\bibfnamefont{R.}~\bibnamefont{Santra}} \bibnamefont{and}
  \bibinfo{author}{\bibfnamefont{L.~S.} \bibnamefont{Cederbaum}},
  \bibinfo{journal}{Phys. Rep.} \textbf{\bibinfo{volume}{368}},
  \bibinfo{pages}{1} (\bibinfo{year}{2002}).

\bibitem[{\citenamefont{Siegert}(1939)}]{Siegert:DF-39}
\bibinfo{author}{\bibfnamefont{A.~J.~F.} \bibnamefont{Siegert}},
  \bibinfo{journal}{Phys. Rev.} \textbf{\bibinfo{volume}{56}},
  \bibinfo{pages}{750} (\bibinfo{year}{1939}).

\bibitem[{\citenamefont{Schafer}(2008)}]{Schafer:NM-08}
\bibinfo{author}{\bibfnamefont{K.~J.} \bibnamefont{Schafer}}, in
  \emph{\bibinfo{booktitle}{Strong field laser physics}}, edited by
  \bibinfo{editor}{\bibfnamefont{T.}~\bibnamefont{Brabec}}
  (\bibinfo{publisher}{Springer}, \bibinfo{address}{New York},
  \bibinfo{year}{2008}), vol. \bibinfo{volume}{134} of
  \emph{\bibinfo{series}{Springer series in optical sciences}}, pp.
  \bibinfo{pages}{111--145}, ISBN \bibinfo{isbn}{978-0-378-40077-8}.

\bibitem[{\citenamefont{Craig and Thirunamachandran}(1984)}]{Craig:MQ-84}
\bibinfo{author}{\bibfnamefont{D.~P.} \bibnamefont{Craig}} \bibnamefont{and}
  \bibinfo{author}{\bibfnamefont{T.}~\bibnamefont{Thirunamachandran}},
  \emph{\bibinfo{title}{Molecular quantum electrodynamics}}
  (\bibinfo{publisher}{Academic Press}, \bibinfo{address}{London},
  \bibinfo{year}{1984}), ISBN \bibinfo{isbn}{0-486-40214-2}.

\bibitem[{\citenamefont{Barth and Lasser}(2009)}]{Barth:TP-09}
\bibinfo{author}{\bibfnamefont{I.}~\bibnamefont{Barth}} \bibnamefont{and}
  \bibinfo{author}{\bibfnamefont{C.}~\bibnamefont{Lasser}},
  \bibinfo{journal}{J. Phys. B} \textbf{\bibinfo{volume}{42}},
  \bibinfo{pages}{235101} (\bibinfo{year}{2009}).

\bibitem[{\citenamefont{Arfken and Weber}(2005)}]{Arfken:MM-05}
\bibinfo{author}{\bibfnamefont{G.~B.} \bibnamefont{Arfken}} \bibnamefont{and}
  \bibinfo{author}{\bibfnamefont{H.~J.} \bibnamefont{Weber}},
  \emph{\bibinfo{title}{Mathematical methods for physicists}}
  (\bibinfo{publisher}{Elsevier Academic Press}, \bibinfo{address}{New York},
  \bibinfo{year}{2005}), \bibinfo{edition}{sixth} ed.

\bibitem[{\citenamefont{Brabec and Krausz}(2000)}]{Brabec:IF-00}
\bibinfo{author}{\bibfnamefont{T.}~\bibnamefont{Brabec}} \bibnamefont{and}
  \bibinfo{author}{\bibfnamefont{F.}~\bibnamefont{Krausz}},
  \bibinfo{journal}{Rev. Mod. Phys.} \textbf{\bibinfo{volume}{72}},
  \bibinfo{pages}{545} (\bibinfo{year}{2000}).

\bibitem[{\citenamefont{Gaarde et~al.}(2008)\citenamefont{Gaarde, Tate, and
  Schafer}}]{Gaarde:MA-08}
\bibinfo{author}{\bibfnamefont{M.~B.} \bibnamefont{Gaarde}},
  \bibinfo{author}{\bibfnamefont{J.~L.} \bibnamefont{Tate}}, \bibnamefont{and}
  \bibinfo{author}{\bibfnamefont{K.~J.} \bibnamefont{Schafer}},
  \bibinfo{journal}{J. Phys. B} \textbf{\bibinfo{volume}{41}},
  \bibinfo{pages}{132001} (\bibinfo{year}{2008}).

\bibitem[{\citenamefont{L'Huillier et~al.}(1992)\citenamefont{L'Huillier,
  Balcou, Candel, Schafer, and Kulander}}]{L'Hu922778}
\bibinfo{author}{\bibfnamefont{A.}~\bibnamefont{L'Huillier}},
  \bibinfo{author}{\bibfnamefont{P.}~\bibnamefont{Balcou}},
  \bibinfo{author}{\bibfnamefont{S.}~\bibnamefont{Candel}},
  \bibinfo{author}{\bibfnamefont{K.~J.} \bibnamefont{Schafer}},
  \bibnamefont{and} \bibinfo{author}{\bibfnamefont{K.~C.}
  \bibnamefont{Kulander}}, \bibinfo{journal}{Phys. Rev. A}
  \textbf{\bibinfo{volume}{46}}, \bibinfo{pages}{2778} (\bibinfo{year}{1992}).

\bibitem[{\citenamefont{Priori et~al.}(2000)\citenamefont{Priori, Cerullo,
  Nisoli, Stagira, Silvestri, Villoresi, Poletto, Ceccherini, Altucci, Bruzzese
  et~al.}}]{Priori2000}
\bibinfo{author}{\bibfnamefont{E.}~\bibnamefont{Priori}},
  \bibinfo{author}{\bibfnamefont{G.}~\bibnamefont{Cerullo}},
  \bibinfo{author}{\bibfnamefont{M.}~\bibnamefont{Nisoli}},
  \bibinfo{author}{\bibfnamefont{S.}~\bibnamefont{Stagira}},
  \bibinfo{author}{\bibfnamefont{S.~D.} \bibnamefont{Silvestri}},
  \bibinfo{author}{\bibfnamefont{P.}~\bibnamefont{Villoresi}},
  \bibinfo{author}{\bibfnamefont{L.}~\bibnamefont{Poletto}},
  \bibinfo{author}{\bibfnamefont{P.}~\bibnamefont{Ceccherini}},
  \bibinfo{author}{\bibfnamefont{C.}~\bibnamefont{Altucci}},
  \bibinfo{author}{\bibfnamefont{R.}~\bibnamefont{Bruzzese}},
  \bibnamefont{et~al.}, \bibinfo{journal}{Phys. Rev. A}
  \textbf{\bibinfo{volume}{61}}, \bibinfo{pages}{063801}
  (\bibinfo{year}{2000}).

\bibitem[{\citenamefont{Yakovlev et~al.}(2007)\citenamefont{Yakovlev, Ivanov,
  and Krausz}}]{Yak0715351}
\bibinfo{author}{\bibfnamefont{V.~S.} \bibnamefont{Yakovlev}},
  \bibinfo{author}{\bibfnamefont{M.~Y.} \bibnamefont{Ivanov}},
  \bibnamefont{and} \bibinfo{author}{\bibfnamefont{F.}~\bibnamefont{Krausz}},
  \bibinfo{journal}{Opt. Express} \textbf{\bibinfo{volume}{15}},
  \bibinfo{pages}{15351} (\bibinfo{year}{2007}).

\bibitem[{\citenamefont{Buth and
  Santra}(2008{\natexlab{b}})}]{fella:pgm-V1.3.0}
\bibinfo{author}{\bibfnamefont{C.}~\bibnamefont{Buth}} \bibnamefont{and}
  \bibinfo{author}{\bibfnamefont{R.}~\bibnamefont{Santra}},
  \emph{\bibinfo{title}{\textsc{fella}~-- the free electron laser atomic,
  molecular, and optical physics program package}},
  \bibinfo{organization}{Argonne National Laboratory},
  \bibinfo{address}{Argonne, Illinois, USA}
  (\bibinfo{year}{2008}{\natexlab{b}}), \bibinfo{note}{version~1.3.0, with
  contributions by Mark Baertschy, Kevin Christ, Chris H.{} Greene, Hans-Dieter
  Meyer, and Thomas Sommerfeld,
  \href{http://chemistry.anl.gov/fundamental_interactions/AMO_fella.html}
  {chemistry.anl.gov/Fundamental\_Interactions/FELLA\_main.shtml}}.

\bibitem[{\citenamefont{Koonin}(1998)}]{Koonin98}
\bibinfo{author}{\bibfnamefont{S.~E.} \bibnamefont{Koonin}},
  \emph{\bibinfo{title}{Computational Physics}} (\bibinfo{publisher}{Westview
  Press}, \bibinfo{address}{Boulder, CO}, \bibinfo{year}{1998}).

\bibitem[{\citenamefont{Merzbacher}(1998)}]{Merzbacher:QM-98}
\bibinfo{author}{\bibfnamefont{E.}~\bibnamefont{Merzbacher}},
  \emph{\bibinfo{title}{Quantum mechanics}} (\bibinfo{publisher}{John Wiley~\&
  Sons}, \bibinfo{address}{New York}, \bibinfo{year}{1998}),
  \bibinfo{edition}{3rd} ed., ISBN \bibinfo{isbn}{0-471-88702-1}.

\bibitem[{\citenamefont{Moiseyev}(1998{\natexlab{b}})}]{Moiseyev:DU-98}
\bibinfo{author}{\bibfnamefont{N.}~\bibnamefont{Moiseyev}},
  \bibinfo{journal}{J. Phys. B} \textbf{\bibinfo{volume}{31}},
  \bibinfo{pages}{1431} (\bibinfo{year}{1998}{\natexlab{b}}).

\bibitem[{\citenamefont{Riss and Meyer}(1998)}]{Riss:TC-98}
\bibinfo{author}{\bibfnamefont{U.~V.} \bibnamefont{Riss}} \bibnamefont{and}
  \bibinfo{author}{\bibfnamefont{H.-D.} \bibnamefont{Meyer}},
  \bibinfo{journal}{J. Phys. B} \textbf{\bibinfo{volume}{31}},
  \bibinfo{pages}{2279} (\bibinfo{year}{1998}).

\bibitem[{\citenamefont{Karlsson}(1998)}]{Karlsson:AR-98}
\bibinfo{author}{\bibfnamefont{H.~O.} \bibnamefont{Karlsson}},
  \bibinfo{journal}{J. Chem. Phys.} \textbf{\bibinfo{volume}{109}},
  \bibinfo{pages}{9366} (\bibinfo{year}{1998}).

\bibitem[{\citenamefont{Diels and Rudolph}(2006)}]{Diels:UL-06}
\bibinfo{author}{\bibfnamefont{J.-C.} \bibnamefont{Diels}} \bibnamefont{and}
  \bibinfo{author}{\bibfnamefont{W.}~\bibnamefont{Rudolph}},
  \emph{\bibinfo{title}{Ultrashort laser pulse phenomena}}, Optics and
  photonics series (\bibinfo{publisher}{Academic Press},
  \bibinfo{address}{Amsterdam}, \bibinfo{year}{2006}), \bibinfo{edition}{2nd}
  ed., ISBN \bibinfo{isbn}{0-12-215493-2}.

\bibitem[{\citenamefont{Henke et~al.}(1993)\citenamefont{Henke, Gullikson, and
  Davis}}]{Hen93181}
\bibinfo{author}{\bibfnamefont{B.}~\bibnamefont{Henke}},
  \bibinfo{author}{\bibfnamefont{E.}~\bibnamefont{Gullikson}},
  \bibnamefont{and} \bibinfo{author}{\bibfnamefont{J.}~\bibnamefont{Davis}},
  \bibinfo{journal}{Atomic Data and Nuclear Data Tables}
  \textbf{\bibinfo{volume}{54}}, \bibinfo{pages}{181} (\bibinfo{year}{1993}).

\bibitem[{\citenamefont{Yost et~al.}(2009)\citenamefont{Yost, Schibli, Ye,
  Tate, Hostetter, Schafer, and Gaarde}}]{Yost:BT-09}
\bibinfo{author}{\bibfnamefont{D.~C.} \bibnamefont{Yost}},
  \bibinfo{author}{\bibfnamefont{T.~R.} \bibnamefont{Schibli}},
  \bibinfo{author}{\bibfnamefont{J.}~\bibnamefont{Ye}},
  \bibinfo{author}{\bibfnamefont{J.~L.} \bibnamefont{Tate}},
  \bibinfo{author}{\bibfnamefont{J.}~\bibnamefont{Hostetter}},
  \bibinfo{author}{\bibfnamefont{K.~J.} \bibnamefont{Schafer}},
  \bibnamefont{and} \bibinfo{author}{\bibfnamefont{M.~B.}
  \bibnamefont{Gaarde}}, \bibinfo{journal}{Nat. Phys.}
  \textbf{\bibinfo{volume}{5}}, \bibinfo{pages}{815} (\bibinfo{year}{2009}).

\bibitem[{\citenamefont{He et~al.}(2009)\citenamefont{He, Miranda, Schwenke,
  Guilbaud, Ruchon, Heyl, Georgiadou, Rakowski, Persson, Gaarde
  et~al.}}]{He09063829}
\bibinfo{author}{\bibfnamefont{X.}~\bibnamefont{He}},
  \bibinfo{author}{\bibfnamefont{M.}~\bibnamefont{Miranda}},
  \bibinfo{author}{\bibfnamefont{J.}~\bibnamefont{Schwenke}},
  \bibinfo{author}{\bibfnamefont{O.}~\bibnamefont{Guilbaud}},
  \bibinfo{author}{\bibfnamefont{T.}~\bibnamefont{Ruchon}},
  \bibinfo{author}{\bibfnamefont{C.}~\bibnamefont{Heyl}},
  \bibinfo{author}{\bibfnamefont{E.}~\bibnamefont{Georgiadou}},
  \bibinfo{author}{\bibfnamefont{R.}~\bibnamefont{Rakowski}},
  \bibinfo{author}{\bibfnamefont{A.}~\bibnamefont{Persson}},
  \bibinfo{author}{\bibfnamefont{M.~B.} \bibnamefont{Gaarde}},
  \bibnamefont{et~al.}, \bibinfo{journal}{Phys. Rev. A}
  \textbf{\bibinfo{volume}{79}}, \bibinfo{pages}{063829}
  (\bibinfo{year}{2009}).
\end{thebibliography}
\end{document}